\documentclass[onecolumn,secnumarabic,amsmath,amssymb,balancelastpage,nofootinbib]{article}

\usepackage[e]{esvect}
\usepackage{xcolor}
\usepackage{graphics}      
\usepackage{graphicx}      
\usepackage{epsf}          
\usepackage{bm}            

\usepackage{natbib}
\usepackage{amssymb}
\usepackage{amsmath,esint}
\usepackage{mathrsfs}
\usepackage{framed}
\usepackage{bigints} 
\usepackage{enumitem}
\usepackage{soul} 
\usepackage{pifont}
\usepackage[capbesideposition={left,center},facing=yes,capbesidewidth=8cm,capbesidesep=quad]{floatrow} 
\usepackage{setspace} 

\usepackage{pdfcolfoot} 

\usepackage[none]{hyphenat} 

\usepackage[colorlinks=true]{hyperref}  

\usepackage[normalem]{ulem}

\usepackage{epigraph}

\definecolor{darkred}{rgb}{0.6,0,0}
\definecolor{darkgreen}{rgb}{0,0.5,0}
\definecolor{darkblue}{rgb}{0,0,0.6}
\definecolor{purple}{rgb}{0.5,0,0.6}
\hypersetup{ colorlinks,
linkcolor=darkblue,
filecolor=darkgreen,
urlcolor=darkgreen,
citecolor=darkred }

\begin{document}

\sloppy 

\bibliographystyle{authordate1}

\title{\Huge{How do Laws Produce the Future?}}
\author{Charles T. Sebens\\Division of the Humanities and Social Sciences\\California Institute of Technology}
\date{arXiv v.3\ \ \ \ January 4, 2025\vspace*{10 pt}\\Forthcoming in \textit{Philosophy of Physics}}

\maketitle
\vspace*{0 pt}
\begin{abstract}
The view that the laws of nature produce later states of the universe from earlier ones (prominently defended by Maudlin) faces difficult questions as to how the laws produce the future and whether that production is compatible with special relativity.  This article grapples with those questions, arguing that the concerns can be overcome through a close analysis of the laws of classical mechanics and electromagnetism.  The view that laws produce the future seems to require that the laws of nature take a certain form, fitting what Adlam has called ``the time evolution paradigm.''  Making that paradigm precise, we might demand that there be temporally local dynamical laws that take properties of the present and the arbitrarily-short past as input, returning as output changes in such properties into the arbitrarily-short future.  In classical mechanics, Newton's second law can be fit into this form if we follow a proposal from Easwaran and understand the acceleration that appears in the law to capture how velocity (taken to be a property of the present and the arbitrarily-short past) changes into the arbitrarily-short future.  The dynamical laws of electromagnetism can be fit into this form as well, though because electromagnetism is a special relativistic theory we might require that the laws meet a higher standard: linking past light-cone to future light-cone.  With some work, the laws governing the evolution of the vector and scalar potentials in the Lorenz gauge, as well as the evolution of charged matter, can be put in a form that meets this higher standard.
\end{abstract}

\newpage
\tableofcontents
\newpage

\section{Introduction}\label{INTROsection}

In physics, we search for laws of nature.  These laws are often described as governing the evolution of systems over time---producing, or generating, later states from earlier ones.  As \citet[pg.\ 182]{maudlin2007} puts it:
\begin{quote}
``The universe, as well as all the smaller parts of it, is \emph{made}: it is an ongoing enterprise, generated from a beginning and guided towards its future by physical law.''
\end{quote}
This dynamic production picture fits well with the way that many physical theories are taught and understood.  However, we will see here that it is not as easy as it might seem to work out the technical details, even for two parts of physics that seem particularly amenable to a dynamic production interpretation: Newtonian gravity and electromagnetism.  To properly assess the idea that the laws of nature produce the future and to understand its implications, we must determine whether the intuitive picture can be made precise.

The dynamic production account of physical laws can be contrasted with two alternatives.\footnote{There are other options.  \citet{adlam2022}, \citet{chengoldstein}, and \citet{chen2023} discuss the universals account (Platonic reductionism), the powers account (Aristotelian reductionism), and the modal structure account.}  The first is the Humean Mill-Ramsey-Lewis best systems account, according to which the laws do not truly govern what happens in nature, but instead merely offer concise and informative descriptions of all the things that happen throughout the entire history of the universe (including its full past and future).\footnote{See \citet{loewer1996, lange2008}.}  Put poetically, the laws are patterns in the mosaic that is our universe.  The second is a recent alternative, put forward by \citet{adlam2022},  \citet{chengoldstein}, and \citet{meacham2022, meachamF}, according to which the laws govern without dynamic production by placing constraints on possible histories for the universe (constraints that may or may not look like laws of time evolution).

There is much that could be said about the advantages and disadvantages of these three competing accounts.  Here, my focus will be on developing the dynamic production and defending it from two serious challenges.  First, one might wonder how the laws produce the future.  Second, one might be worried as to whether that production is compatible with special relativity.

Let us begin with the first challenge.  Asking how the laws produce the future verges on the unanswerable, as the dynamic production account takes the action of the laws to be primitive---not something that can be analyzed in terms of anything more fundamental.  But, there is a sensible question here.  \citet[pg.\ 60]{chengoldstein} and \citet{dorst} ask about the relata of the production relation: What is being produced by what?  If we take the input of the laws to be the state of the world at just one moment, we only have a static snapshot that leaves out features of the world---like velocities---that are needed to determine future evolution (\citealp[pg.\ 46 \& 60]{chengoldstein}).  Specifying the output is also tricky.  The output of the laws cannot be the state of the world at the next moment because (if time is continuous) there is no next moment \citep[pg.\ 133]{loewer2012}.

For the dynamic production account to be viable, the laws of nature must take a certain form.  They must include dynamical laws that specify (either deterministically or probabilistically) how the world evolves from one moment to the next.  The laws must fit what \citet{adlam2022, adlam2022c} has called ``the time evolution paradigm.''\footnote{\citet{smolin2009} and \citet{wharton2014, wharton2015} earlier called this the ``Newtonian schema.''}  Making that paradigm precise, we might demand that there be temporally local dynamical laws that take properties of the present and the arbitrarily-short past as input and return as output changes in such properties into the arbitrarily-short future.  For theories like this, we have a clear picture as to how the laws produce the future and can respond to the first challenge.  The input to the laws is not just the present moment and the output is not the state at any particular future moment.

Temporally local dynamical laws are not hard to find.  In classical mechanics, Newton's second law can be fit into the above form if we follow a proposal from \citet{easwaran2014} and understand the acceleration that appears in the law to capture how velocity (taken to be a property of the present and the arbitrarily-short past) changes into the arbitrarily-short future.  Classical electromagnetism can also be formulated as a theory with temporally local dynamical laws, though there exist alternative formulations that are certainly not temporally local (the Wheeler-Feynman and retarded action-at-a-distance approaches).

Let us now turn to the second challenge.  \citet{dorst} has criticized the dynamic production account for being in tension with the special theory of relativity.  One objection is that special relativity pushes us to a block universe theory where the past and future are just as real as the present and there is no work for the laws to do in producing the future.  The future is already there.  This objection will only be addressed briefly in section \ref{TEPsection}.  Dorst's other objection is that we cannot say that dynamic production is occurring from one time slice to the next without privileging a particular way of carving the universe into simultaneity slices.  Maudlin is willing to introduce such a preferred foliation because he believes that, in quantum physics, a preferred foliation is the most natural way to account for the experiments that illustrate Bell's theorem.  Maudlin may be right about how the quantum physics will shake out, but I still think that it is valuable to understand how relativistic dynamic production might work without a preferred foliation for two reasons.  First, there are important classical theories that do not need a preferred foliation.  Second, quantum theories that include parallel universes may not need such a foliation.

In a relativistic theory like classical electromagnetism, we can go beyond merely requiring temporally local dynamical laws that connect past and future and require that the dynamical laws be spatiotemporally local, connecting the past and future light-cones at every space-time point.  This higher standard, which can be called ``the relativistic time evolution paradigm,'' supports an understanding of dynamic production that does not require a preferred foliation.

The article proceeds as follows:  The next section presents a precise statement of the time evolution paradigm that theories must fit for the dynamic production account of laws to be tenable.  Section \ref{CMsection} explores multiple ways of fitting Newton's classical theory of gravity into the time evolution paradigm, highlighting some challenges and choice points that will be relevant to the subsequent discussion of electromagnetism.  Section \ref{EMsection1} gives a simple way to cast electromagnetism into the time evolution paradigm.  Section \ref{RTEPsection} puts forward the relativistic time evolution paradigm.  Section \ref{EMsection2} explores (in depth) how the laws of electromagnetism might be fit into the relativistic time evolution paradigm, showing that it is possible and leaving open whether there might be a better way of doing so.  Section \ref{QFTGRsection} looks ahead to more advanced physics and considers the prospects for interpreting quantum field theory and general relativity as theories of dynamic production.

\section{The Time Evolution Paradigm}\label{TEPsection}

The idea that laws of nature dictate how the universe evolves over time may seem old-fashioned.  \citet{adlam2022} writes:
\begin{quote}
``Newton bequeathed to us a picture of physics in which the fundamental role of laws is to give rise to time evolution: the Newtonian universe can be regarded as something like a computer which takes in an initial state and evolves it forward in time \dots\ But science has come a long way since the time of Newton, and thus we should not necessarily expect that accounts of lawhood based on a Newtonian time-evolution picture will be well-suited to the realities of modern physics.'' \citep[pg. 3]{adlam2022} 
\end{quote}
Adlam is not alone in presenting the ``Newtonian time-evolution picture'' as behind the times.  The 2023 Foundational Questions Institute (FQxI) essay contest prompt began with this sentence:
\begin{quote}
``Galileo claimed that the book of nature is written in mathematics, and indeed the discipline that he, Newton, and other 17th-century natural philosophers and mathematicians founded took on a particular form: mathematical laws expressing necessary relations between elements of the world, largely expressed in differential equations governing the time evolution of the state of the world.'' \citep{fqxiprompt}
\end{quote}
The prompt later asks:  ``Could it have been otherwise? Should it be otherwise now?''  The implication is that this is an outdated mold into which physical theories need no longer be cast.  I take the conservative view that this mold should be retained, and that more effort should be put into fitting existing and future theories into the mold.  It has served physics well and should not be abandoned lightly.

In Newton's physics, the laws are deterministic.\footnote{At least, the laws of Newtonian physics are generally presented as deterministic.  The question as to whether the laws truly are deterministic will be raised near the end of this section.}  This is sometimes explained by saying that the laws allow only a single future given just the state at a particular moment, but here we must be careful.  How does one specify the physical state at a moment?  \citet[pg.\ 9--10]{albert2000} gives two requirements for such a specification.  First, the instantaneous state should describe features of that moment alone.  Second, a full set of instantaneous states across all times should fully specify everything that occurs.  You might initially think that the masses, positions, and velocities of bodies should all be included in the instantaneous state, but Albert uses his first requirement to disqualify velocity.  Velocity is the rate at which position changes and the velocity at a moment can only be determined by considering (arbitrarily small) temporal neighborhoods around the moment in question.  Strictly speaking, it is not a property of that moment alone.  Excluding velocity does not lead to a problem with the second requirement because the velocities at any time can be determined once the positions are specified at all times.  If we agree with Albert and say that velocities are not included in instantaneous states, then the laws of Newton's physics will certainly fail to determine a unique future given only the state at a moment.  But, the laws will give a unique future given the state of the world over an arbitrarily short time interval.  Let us thus follow \citet[pg.\ 11]{albert2000} (and \citealp[pg.\ 195]{arntzenius2000}) and say that a theory is \emph{deterministic} if and only if specifying the state of the world over an arbitrarily small time interval to the past of a given moment (assuming the laws to be obeyed during this interval) uniquely determines a single future that is allowed by the laws (a single future that is physically possible).\footnote{\citet{builesF} call this kind of determinism ``Near Markovian Determination.''}

\begin{figure}[htb]
\center{\includegraphics[width=6 cm]{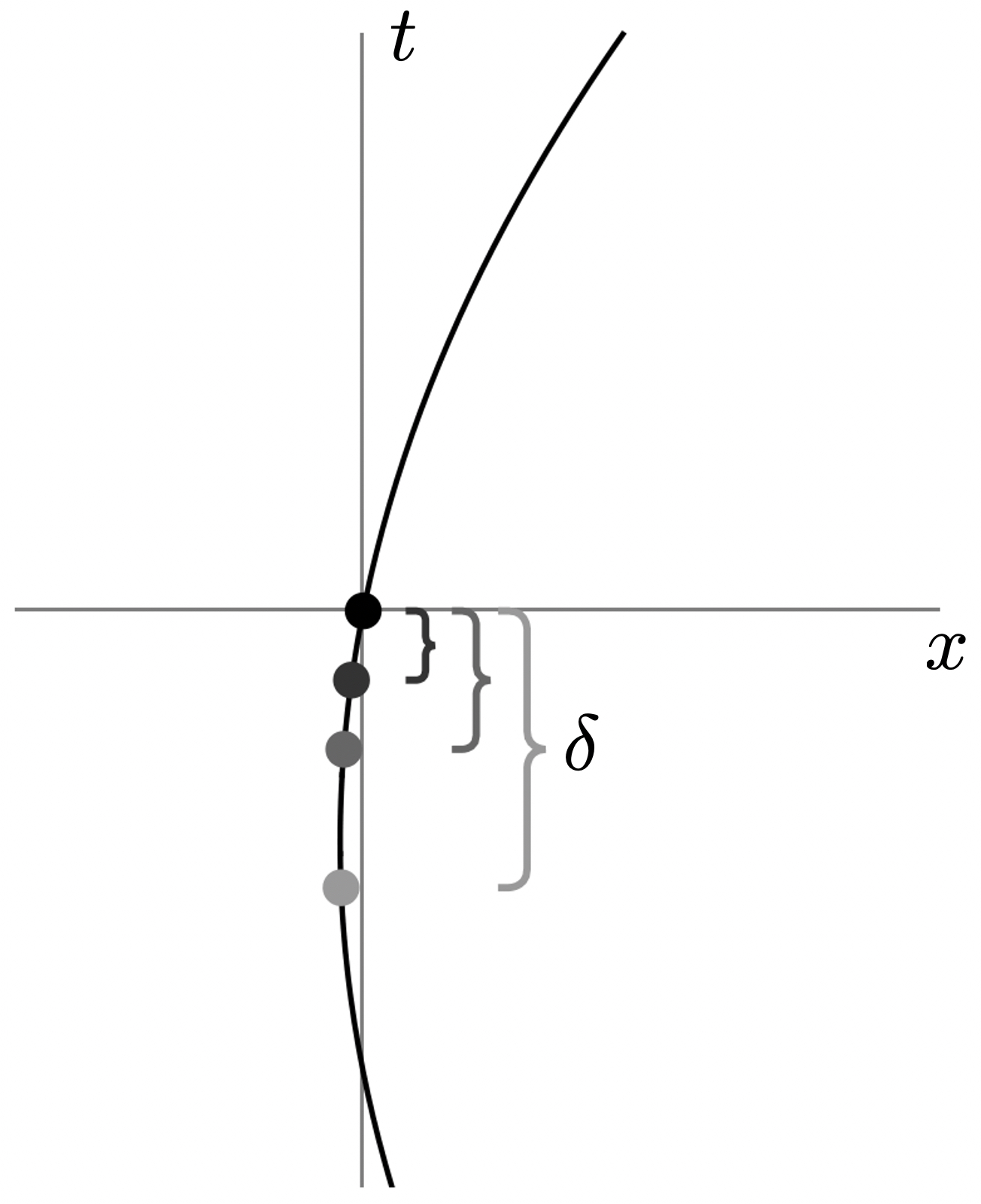}}
\caption{The past velocity $\vec{v}^{\,p}$ \eqref{pastvelocity} of a body is determined by comparing the body's location at a moment to its location at progressively closer past moments (shown here as darkening dots).}
  \label{pastvelocityfig}
\end{figure}

The velocity at a moment can be part of what determines the future of that moment, provided that we understand the relevant velocity to be the past velocity $\vec{v}^{\,p}$: a \emph{past} derivative defined by considering arbitrarily small time intervals to the past of that moment,
\begin{equation}
\vec{v}^{\,p}(t) = \left(\frac{d}{dt}\right)^p \vec{x}(t)= \lim_{\delta \to 0} \frac{\vec{x}(t)-\vec{x}(t-\delta)}{\delta}
\ ,
\label{pastvelocity}
\end{equation}
as in \citet[sec.\ 2]{lange2005}.  This velocity is depicted in figure \ref{pastvelocityfig}.  The future velocity, by contrast, would be defined by considering arbitrarily small time intervals to the future.  In ordinary circumstances, the past and future velocities will agree.  However, they could, in principle, come apart if there is a kink in the trajectory of a body through space and time.  Whether past and future velocities are permitted to come apart will depend on the laws of the theory at hand.

\begin{figure}[htb]
\center{\includegraphics[width=13 cm]{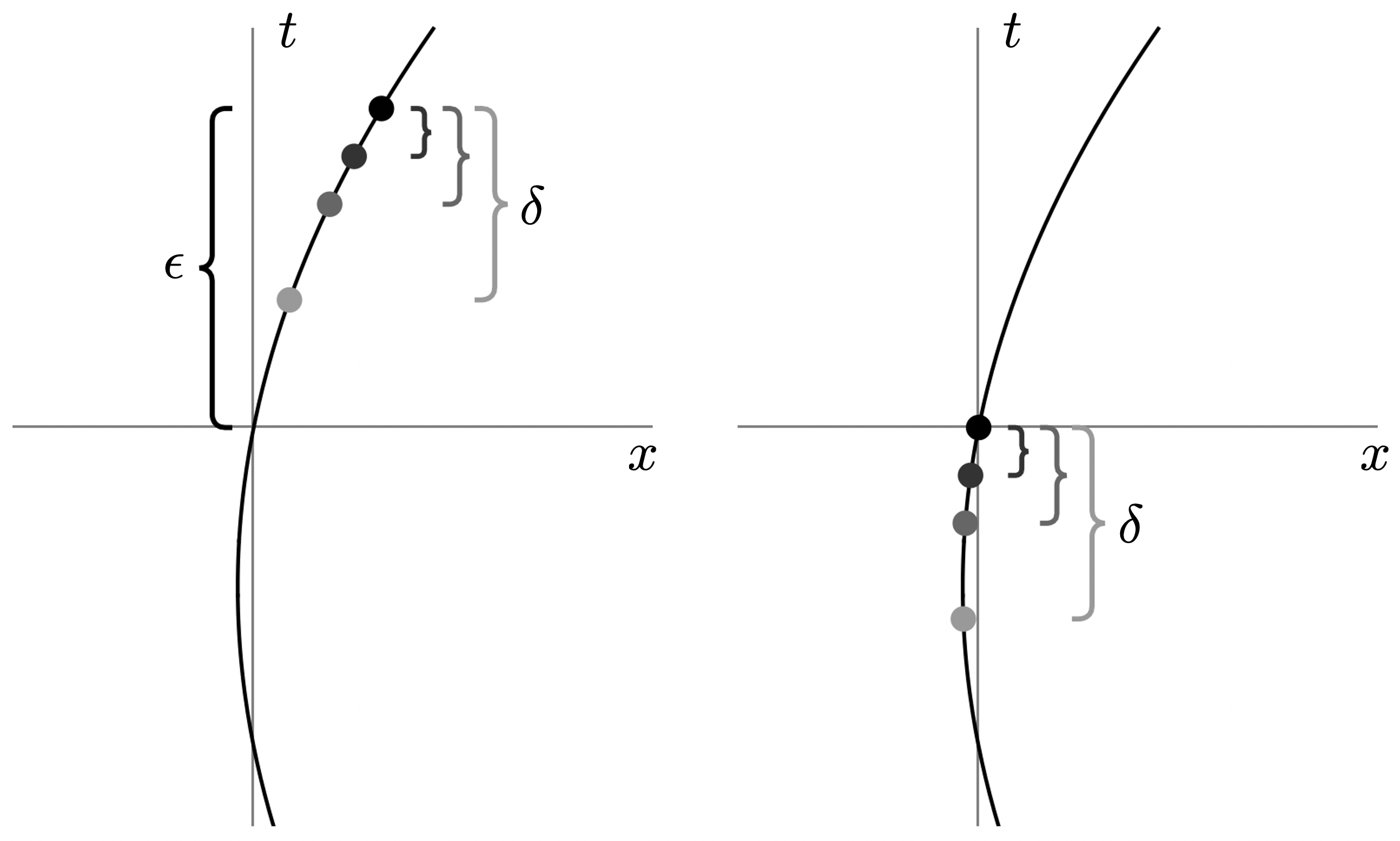}}
\caption{The future acceleration $\vec{a}^{\,pf}$ \eqref{futureacceleration} of a body is determined by comparing (a) the body's past velocity at a time $\epsilon$ in the future (which is determined by looking $\delta$ into the past of that moment, as shown on the left), and (b) the body's past velocity at the moment in question (which is determined by looking $\delta$ into the past of that moment, as shown on the right).  As is evident from the image on the right, a body's future acceleration $\vec{a}^{\,pf}$ depends on its past and is not purely a property of its present and future.}
  \label{futureaccelerationfig}
\end{figure}

Modifying a proposal from \citet{easwaran2014},\footnote{\citet{easwaran2014} uses clever, but more complicated, definitions of the past and future time derivatives.  These definitions avoid referencing the present moment by either considering two arbitrarily close past times (for the past time derivative) or two arbitrarily close future times (for the future derivative).   Easwaran calls these open-ended derivatives (because the endpoint at the present is left out) and contrasts them with the closed-ended derivatives defined above.  Easwaran's choice to use open-ended derivatives allows him to classify past velocity as entirely about the past and future acceleration as entirely about the future.  One reason behind the choice to use closed-ended derivatives here will be given later (in footnote \ref{openended2}) when we see that, in electromagnetism, even if we were to use open-ended derivatives the laws would not give pure properties of the future light-cone as output.\label{openended1}} the acceleration that appears in the laws can be defined as the future acceleration $\vec{a}^{\,pf}$: a \emph{future} derivative of the past velocity,
\begin{equation}
\vec{a}^{\,pf}(t) = \left(\frac{d}{dt}\right)^f \vec{v}^{\,p}(t)= \lim_{\epsilon \to 0} \frac{\vec{v}^{\,p}(t+\epsilon)-\vec{v}^{\,p}(t)}{\epsilon}
\ .
\label{futureacceleration}
\end{equation}
Thus, in Newton's second law the acceleration of a body can be understood as a future effect of present forces: $\vec{F}=m\vec{a}^{\,pf}$.  This acceleration is depicted in figure \ref{futureaccelerationfig}.

The velocity in \eqref{pastvelocity} is an example of a past neighborhood property:\footnote{Here I have modified the definition of past neighborhood property in \citep[pg.\ 849]{easwaran2014}.  The square brackets indicate that the endpoints of the intervals are included.}
\begin{quote}
A \textbf{past neighborhood property} of an object at $t$ is not determined by the present properties of the object at $t$ alone, but is determined once the present properties of the object are specified within any arbitrarily small interval $[ t-\Delta,t ]$.\footnote{Because we need at least some of the past to determine the velocity, but we do not need any particular past moment, there is a puzzle as to exactly what are the facts about position in virtue of which the velocity at a moment is what it is \citep{builes2019}.}
\end{quote}
A future neighborhood property could be defined similarly, replacing $[ t-\Delta,t ]$ by $[ t,t+\Delta ]$.  The acceleration $\vec{a}^{\,pf}$ in \eqref{futureacceleration} can be classified as what I will call a ``past-to-future neighborhood property'': a property that is determined by looking into the arbitrarily-short future and potentially considering within that window both properties of the present (like mass) and the arbitrarily-short past (like velocity).  Put more precisely, 
\begin{quote}
A \textbf{past-to-future neighborhood property} of an object at $t$ is not determined by the present and past neighborhood properties of the object at $t$ alone, but is determined once the present and past neighborhood properties of the object are specified within any arbitrarily small interval $[ t,t+\Delta ]$.
\end{quote}
Note that, by this definition, future neighborhood properties also count as past-to-future neighborhood properties because they are determined by the present properties over an arbitrarily small interval to the future.  However, some past-to-future properties---like the acceleration $\vec{a}^{\,pf}$---are not future neighborhood properties (see figure \ref{futureaccelerationfig}).  If we require dynamical laws to give past-to-future neighborhood properties as output, we can allow for either first-order dynamics (as in the Schr\"{o}dinger equation) where the output is a future neighborhood property (and thus a past-to-future neighborhood property) or second-order dynamics (as in Newton's second law) where the output is a past-to-future neighborhood property (that is not a future neighborhood property).\footnote{One might also wonder about third and higher order dynamics.  \citet{easwaran2014} gives an argument against the possibility of such dynamics (cf.\ \citealp{swanson2022}).}

To avoid action at a temporal distance, \citet[pg.\ 857]{easwaran2014} conjectures that ``The fundamental causal [dynamical] laws must use present properties and past neighbourhood properties to determine future neighbourhood properties.''  This idea has not been widely adopted, but it is an attractive proposal that fits well with a dynamic production account of laws.  Let us take a modified version of the idea onboard here and say that a law counts as a \emph{temporally local dynamical law} if it gives past-to-future neighborhood properties as output and takes as input only present properties or past neighborhood properties.  Dynamical laws that are not temporally local might take as input properties at earlier or later times, as in the retarded action-at-a-distance and Wheeler-Feynman versions of electromagnetism.  Dynamical laws that are temporally local may either take as input only present properties (as in the first-order Schr\"{o}dinger equation) or both present and past neighborhood properties (as in Newton's second law).

Having now set the stage, let us say that a set of laws fits \textbf{The Time Evolution Paradigm} if and only if the following three conditions are met:
\begin{enumerate}
\item \textbf{Temporally Local Dynamical Laws:}  A subset of the laws (the dynamical laws) can be applied at any moment $t$, take as input only present properties or past neighborhood properties, and give as output only past-to-future neighborhood properties.  The dynamical laws may give either precise values for the past-to-future neighborhood properties or probability distributions over such values.
\item \textbf{Non-Dynamical Laws:}  The laws that do not take the above dynamical form (the non-dynamical laws) express relations between present properties or past neighborhood properties that hold at every moment, $t$.
\item \textbf{Deterministic or Stochastic:}  Once a law-abiding history over an arbitrarily small time interval to the past of a given moment has been fixed, the laws either uniquely fix a single future sequence of states or the laws yield a precise probability distribution over future sequences of states.
\end{enumerate}
This formulation allows for non-dynamical laws in addition to dynamical laws.  As a candidate example of a non-dynamical law, \citet[pg.\ 13]{maudlin2007} gives Newton's law of gravitation, which uses present positions and masses to specify the present force on a given body.  Maudlin describes this law as an ``adjunct principle'' that allows the dynamical law $\vec{F}=m\vec{a}$ to predict future time evolution.  \citet[pg.\ 60]{chengoldstein} consider the example of Gauss's law, $\vec{\nabla}\cdot\vec{E}=4 \pi \rho$, which can be regarded as a non-dynamical law relating the present divergence of the electric field to the present distribution of charge.  The requirement in the second condition that non-dynamical laws ``express relations between present properties or past neighborhood properties'' rules out certain putative non-dynamical laws, such as those that posit probability distributions over initial conditions \citep[sec.\ 3.3.3]{chengoldstein}.\footnote{One could still posit a probability distribution over initial conditions to explain thermodynamic asymmetries, as \citet{albert2000} does with his Past Hypothesis and Statistical Postulate.  Within the time evolution paradigm, such a probability distribution would have to be regarded as a hypothesis (or postulate), not a law of nature.}  The third condition, that the laws be either deterministic or stochastic, is meant to exclude laws that give incomplete stories about the future: neither determining the future uniquely nor giving a definite probability distribution over different futures.  As an example of such an incomplete theory, consider a version of electromagnetism where the vector and scalar potentials are taken to be the proper way to represent the electromagnetic field and no gauge fixing condition is imposed.  Then, multiple future evolutions of the potentials will be compatible with the past \citep[pg.\ 14]{maudlin2018}.

The dynamic production account of laws appears to require the laws of nature to fit the time evolution paradigm presented above.  Let us begin with the first condition of the time evolution paradigm.  The temporally local dynamical laws that appear in this condition can serve as what \citet{maudlin2007} calls the fundamental laws of temporal evolution (FLOTEs).\footnote{Relaxing the requirement of temporal locality, one might wonder whether laws that take more of the past as input could plausibly be construed as dynamically producing the future (as in, e.g., a retarded action-at-a-distance formulation of electromagnetism).  Laws like that are indeed compatible with the broad idea of dynamic production, but they do not look like the FLOTEs that Maudlin envisages and I will take them to be at odds with the type of dynamic production account of laws that Maudlin has defended (and that I am defending here).}  It is these laws that generate time evolution, producing future states from earlier ones.  As Maudlin puts it,
\begin{quote}
``The universe started out in some particular initial state. The laws of temporal evolution operate, whether deterministically or stochastically, from that initial state to generate or produce later states.'' \citep[pg.\ 174]{maudlin2007}
\end{quote}
In this quote and elsewhere, it is unclear how central the idea of an initial state is to Maudlin's account.  In any case, the contours of the dynamic production account discussed here need not match Maudlin's exactly.  Let us not build into the account any assumption that there was a first moment.  There may or may not have been.  \citet[pg.\ 61]{chengoldstein} criticize the assumption of a first moment because it would rule out spacetimes without temporal boundaries.  They say that it appears to be an important part of the dynamic production account because ``it is what gets the entire productive enterprise started,'' but I do not see any problem with the idea that production has always been occurring no matter how far back you go in time.  If there were a first moment, it may not be correct to say (as they do) that the first moment gets the productive enterprise started.  For second-order dynamical laws (like Newton's second law), we would need an arbitrarily thin time slice preceding a given moment for the laws to generate time evolution (because we need velocities and other rates of change).  For such laws, any moment after the first can be explained as having been produced by an earlier arbitrarily thin time slice evolving via the laws, but the first moment alone would be insufficient to get the evolution going.

\citet[pg.\ 46 \& 60]{chengoldstein} challenge the idea of dynamic production because the state of the universe at a moment does not contain features like velocity and momentum that are needed to evolve the state forward via the laws.  However, velocity and momentum can be formulated as past neighborhood properties and such properties may serve as inputs to temporally local dynamical laws connecting the present and its past neighborhood to the future.  The way that the dynamical laws in the time evolution paradigm connect past to future also helps to address \citeauthor{loewer2012}'s \citeyearpar[pg.\ 133]{loewer2012} remark that ``there isn't a `next' state if time is continuous'' to support the claim that one state of the universe produces the next via the laws.  The laws give as output a set of past-to-future-neighborhood properties that specify how things change into the future, not the state at the next moment.

The dynamic production account relies on a fundamental asymmetry between past and future (a fundamental arrow of time),\footnote{See \citet{loewer2012}.} but \citet[pg.\ 108--109]{maudlin2007} argues that it is compatible with the block universe theory---according to which the past and future are just as real as the present.  It would be worrisome if the account was not compatible with the block universe theory because special relativity arguably forces us to the block universe theory.\footnote{See \citet{putnam1967, zimmerman2011}.}  \citet{dorst} has challenged the compatibility, asking how the laws could possibly produce the future if the future is already out there.  This is a serious criticism that could be discussed in depth, but let me briefly point out that a similar argument could be given against the possibility of any ordinary act of production within the block universe theory, such as the creation of a sculpture.  How can Michelangelo create the sculpture of David if it is already out there (in the future)?  One way of responding to Dorst would be to argue that just as ordinary acts of production (like Michelangelo's) are compatible with the block universe theory, the dynamic production of the laws is compatible with the block universe theory.  While it is true that from an outside-of-time perspective the future ``already'' exists, production is something that happens within time and at a moment the future will exist but does not \emph{already} exist.

Let us now turn to the second condition of the time evolution paradigm.  Non-dynamical laws play a quite different role from the dynamical laws, constraining the relations between properties of the present and arbitrarily-short past at a moment.  \citet[ch.\ 1]{maudlin2007} calls such laws ``adjunct principles'' and describes them as principles ``that are needed to fill out the FLOTEs in particular contexts, principles about the magnitudes of forces and the form of the Hamiltonian, or about the sorts of physical states that are allowable.''  Following Maudlin,\footnote{\citet{chengoldstein} and \citet{chen2023} portray Maudlin as requiring all laws to be dynamical laws.  I take his written work to leave open the possibility that there may be non-dynamical laws (as is evident in \citealp[pg.\ 13--14]{maudlin2007}; \citealp{maudlin2018}).} I will understand the dynamic production account as allowing for such non-dynamical laws that supplement the dynamical laws.  The non-dynamical laws restrict the starting points from which dynamic production might occur.  Because the non-dynamical laws do not themselves dynamically produce, one might aspire to a more streamlined account where all laws are dynamical laws.  We will see in sections \ref{CMsection} and \ref{EMsection2} that Newton's theory of gravity and classical electromagnetism can be formulated as theories where all of the laws are dynamical laws.

The third condition of the time evolution paradigm requires the laws to be either deterministic or stochastic.  This is a natural requirement for the dynamic production account and it is one that is imposed by \citet[pg.\ 14--20, 174--183]{maudlin2007}.  To dynamically produce the future, the laws must be sufficiently specific that they either yield a unique future or a probability distribution over possible futures.  In a deterministic theory, the dynamical laws must together say enough about past-to-future neighborhood properties to specify a single future evolution---when applied at all times, alongside the non-dynamical laws.  This is the simplest way that the laws might produce the future and a narrow version of the dynamic production account might require the laws to take this form.  By contrast, stochastic laws allow for multiple futures and provide a probability distribution over such futures.\footnote{In the dynamic production account, the probabilities that appear in stochastic laws should be understood as basic propensities that are not to be reduced to anything more fundamental (such as long-run frequencies). See \citet[pg.\ 17--20]{maudlin2007}; \citet{frigg2007}; \citet[pg.\ 118]{loewer2012}.}  Such laws can arguably also be regarded as producing the future, though they do so in a fundamentally random way.  One can write down laws that are neither deterministic nor stochastic, deeming multiple futures to be compatible with a given past.  It is hard to see how such laws could be understood as producing the future.  Such laws do not specify, even in a probabilistic way, which future is to be produced (leaving open multiple possibilities).  Here are three cases where that might occur: First, as was mentioned previously, this kind of indeterminism arises if electromagnetism is formulated in terms of potentials without a gauge fixing condition.  \citet{maudlin2018} advocates adopting a gauge-fixing condition to avoid this indeterminism.  Second, this kind of indeterminism may also occur in general relativity if one allows for the possibility of closed time-like curves \citep{sep-time-machine}.  \citet[pg.\ 175, 189--191]{maudlin2007} forbids closed time-like curves as incompatible with his picture of dynamic production.  Third, there have been worries raised about Newton's laws of classical mechanics leaving open multiple futures and thus not being as deterministic as they appear \citep{earman2004, vanstrien2021, wilson2009}.  A defender of dynamic production would seek sets of laws that avoid such indeterminism.

Unlike the dynamic production account, the best systems account does not require laws to fit the time evolution paradigm.  Still, deterministic or stochastic temporally local dynamical laws have a good shot at being included in the best system because they get you a lot from a little---allowing you to predict the future (at least probabilistically) from an arbitrarily thin time slice.  \citet[ch.\ 7-8]{callender2017} argues that time is a special dimension within the best systems account because the laws that emerge as part of the best system include laws of time evolution.  As \citet[pg.\ 142]{callender2017} puts it, ``Time is that direction on the manifold of events in which we can tell the strongest or most informative stories.''  Although the laws of the best system might include dynamical laws, the role of these laws is different than in the dynamic production account.  The best systems account starts with a history of things that happen and then finds laws as patterns in that history.  By contrast, the dynamic production account views the history as generated by the laws.  The laws are additional features of the universe beyond the things within the universe that are governed by the laws.

A defender of dynamic production may allow that it is metaphysically possible to have laws of nature that violate the time evolution paradigm and cannot be understood as dynamically producing the future from the past.  Adopting this perspective, dynamic production is not part of what it takes for something to be a law.  It is merely a feature of our laws.  It is difficult to say what it takes in general for something to be a law, but you might say that at a minimum it must constrain what is physically possible.  Here the dynamic production account can make peace with constraint accounts \citep{adlam2022, chengoldstein, meacham2022, meachamF}.  Proponents of constraint accounts prefer the flexibility of their picture, arguing that the laws of our world may well turn out not to include dynamical laws (or to otherwise violate the time evolution paradigm) and that our philosophical theorizing should not preclude this possibility.\footnote{In addition to \citet{chengoldstein, adlam2022, meacham2022, meachamF}, see also \citet{wharton2015, lam2023}.}  Proponents of dynamic production stick their necks out, betting that the final laws will fit the time evolution paradigm because our most successful extant theories are theories of time evolution (or at least can be formulated as theories of time evolution).  This debate hinges primarily on difficult questions about the interpretation of general relativity and quantum field theory that will only be discussed briefly in section \ref{QFTGRsection}.  Here we focus on theories that seem like they should fit the time evolution paradigm, to better understand how dynamic production works in these cases.

\section{Newtonian Gravity}\label{CMsection}

In classical mechanics with point particles, you might say that there is just one dynamical law: Newton's second law (force equals mass times acceleration).  Following \citet{easwaran2014}, as discussed in section \ref{TEPsection}, we can understand this law as taking the net force $\vec{F}_i$ on a given mass $m_i$ as input and giving the body's future acceleration \eqref{futureacceleration} as output,
\begin{equation}
\vec{F}_i=m_i\vec{a}^{\,pf}_i\quad \mbox{(dynamical law)}
\ .
\label{secondlaw}
\end{equation}
For this to be a dynamical law fitting the time evolution paradigm, the force would have to be a present or past neighborhood property.  If the only forces at play are gravitational forces, then the net force on a mass is a present property set by Newton's law of gravitation,
\begin{equation}
\vec{F}_i = \sum_{j \neq i}- G \frac{m_i m_j}{r_{ji}^2}\hat{r}_{ji}\quad \mbox{(non-dynamical law)}
\ .
\label{lawofgravitation}
\end{equation}
Here $\hat{r}_{ji}$ is the unit vector pointing from $m_j$ to $m_i$ and $r_{ji}$ is the distance between $m_j$ and $m_i$.  \citet[pg.\ 13]{maudlin2007} describes Newton's law of gravitation as a non-dynamical law (an ``adjunct principle'').  Including it as such would yield a set of laws that fit the time evolution paradigm: \eqref{secondlaw} and \eqref{lawofgravitation}.  In this case, it is straightforward to combine the dynamical and non-dynamical laws to get a single dynamical law that can govern on its own (without the need for any non-dynamical laws),
\begin{equation}
\vec{a}^{\,pf}_i=\sum_{j \neq i}- G \frac{m_j}{r_{ji}^2}\hat{r}_{ji}\quad \mbox{(dynamical law)}
\ .
\label{combolaw}
\end{equation}
We can thus formulate Newtonian gravity either as a theory with two laws (one dynamical and one not), \eqref{secondlaw} and \eqref{lawofgravitation}, or as a theory with a single dynamical law, \eqref{combolaw}.

One might object that Newtonian gravity is supposed to be a time-symmetric theory, but the earlier characterization of Newton's second law in terms of future acceleration \eqref{secondlaw} is time-asymmetric.  That law tells you how velocities change towards the future, not the past.  A couple remarks on this asymmetry:  First, although standard presentations of the equations of Newtonian gravity may not display any such time-asymmetry, explanations of those equations are generally time-asymmetric.  Forces are described as causing future acceleration, an understanding that fits well with \eqref{secondlaw} \citep[pg.\ 852--853]{easwaran2014}.  Second, the theory remains time-symmetric in the sense that it is time-reversal invariant: for any history that obeys \eqref{secondlaw} and \eqref{lawofgravitation}, the time-reversed history will obey \eqref{secondlaw} and \eqref{lawofgravitation}.  The key point here is that any history where $\vec{F}=m\vec{a}^{\,pf}$ is also a history where $\vec{F}=m\vec{a}^{\,fp}$ because $\vec{a}^{\,pf}$ cannot differ from $\vec{a}^{\,fp}$ if particles are only interacting through well-behaved gravitational forces.

The above story about the laws of Newtonian gravity changes if you take the ontology (what exists according to the theory) to include a gravitational field $\vec{g}$ in addition to the point masses.  To include the field, you might try the following simple (but ultimately unsatisfactory) collection of laws:
\begin{align}
\vec{F}_i=m_i\vec{a}^{\,pf}_i&\quad \mbox{(dynamical law)}
\nonumber
\\
\vec{F}_i = m_i \vec{g}(\vec{x}_i) &\quad \mbox{(non-dynamical law)}
\nonumber
\\
\vec{g}(\vec{x}) = \sum_{j}- G \frac{m_j}{|\vec{x}-\vec{x}_j|^2} &\quad \mbox{(non-dynamical law)}
 \ .
 \label{gfieldlaws}
\end{align}
The first law is Newton's second law, as in \eqref{secondlaw}.  The second law relates the force on a body to the gravitational field at the body's location $\vec{x}_i$.  The third law gives the gravitational field resulting from a certain arrangement of masses as a sum of contributions from each mass.  It is interesting to note that the gravitational field does not have its evolution determined by a dynamical law.  It is rebuilt at each moment by a non-dynamical law.

\begin{figure}[htb]
\center{\includegraphics[width=5 cm]{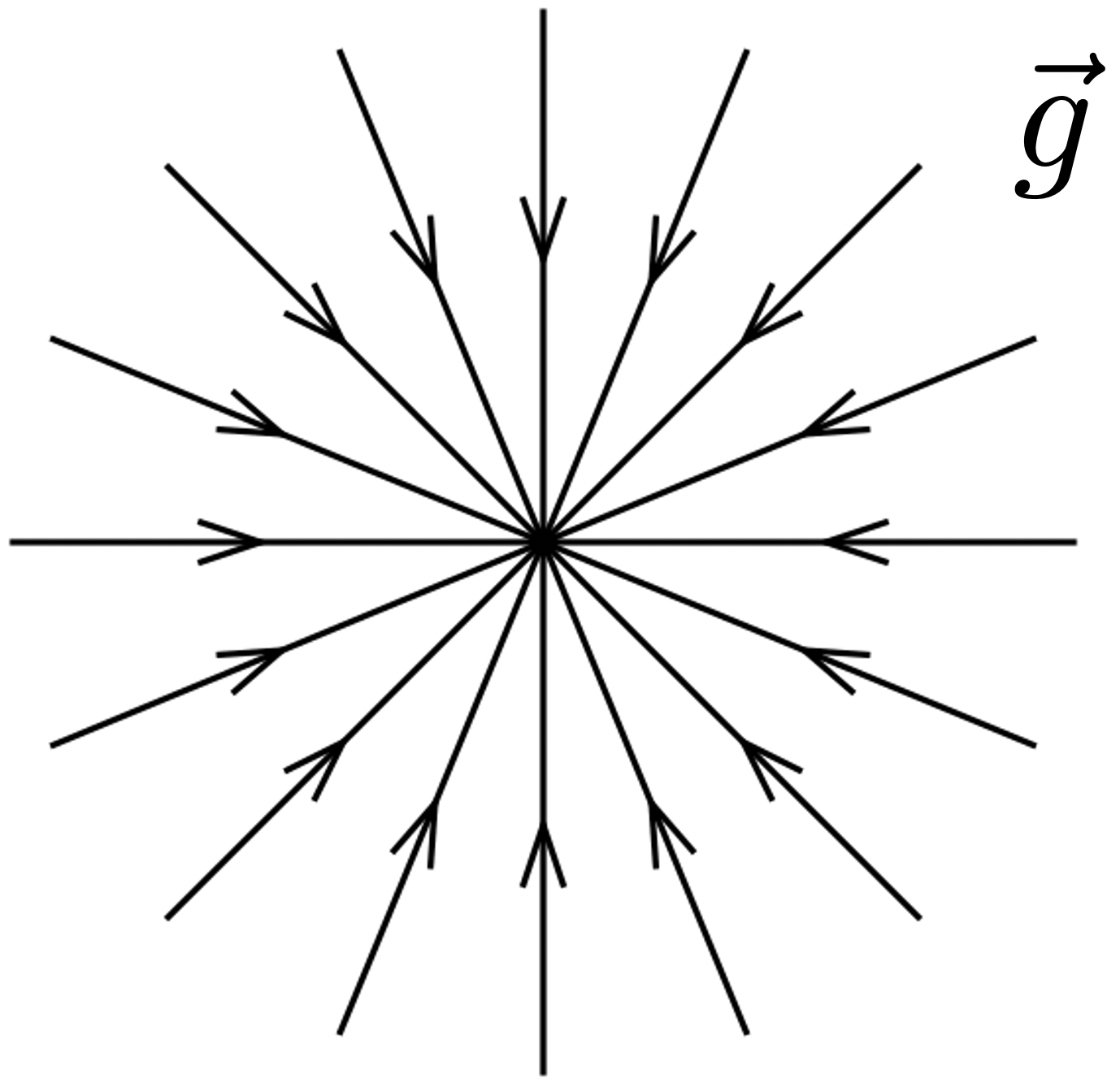}}
\caption{The gravitational field of a point mass is ill-defined at the location of that point mass.}
  \label{gfig}
\end{figure}

The set of laws for Newtonian gravity (including a gravitational field) that were just given \eqref{gfieldlaws} will not function properly because they have problems with self-interaction.  As you approach the location of any point mass, the gravitational field becomes infinitely strong and does not have a clear direction (figure \ref{gfig}). The field is ill-defined at the mass's location, which is exactly where you would like to use it to calculate the force on the mass.  One way to address this problem is to introduce a separate gravitational field sourced by each particle and stipulate that particles do not feel their own fields.  Then, you could derive Newton's law of gravitation \eqref{lawofgravitation} with the important $j \neq i$ in the sum.

A different strategy for addressing self-interaction would be to retain a single gravitational field and replace point masses by extended bodies, either rigid bodies or continua.\footnote{For discussion of the different possible ontologies for classical mechanics (point particles, rigid bodies, and continua), see \citet{wilson1998, wilson2013, vanstrien2021}.}  Then, we would have laws like:
\begin{align}
\vec{\nabla}\cdot\vec{g}=-4\pi G \rho^m &\quad \mbox{(non-dynamical law)}
\\
\vec{\nabla}\times\vec{g}=0 &\quad \mbox{(non-dynamical law)}
\\
\vec{f} = \rho^m \vec{g} &\quad \mbox{(non-dynamical law)}
\ ,
\end{align}
where $G$ is the gravitational constant, $\rho^m$ is the mass density, and $\vec{f}$ is the density of force exerted by the gravitational field on matter.  (Note that because Newtonian gravity closely parallels electrostatics, the first two laws resemble Maxwell's equations as they apply to electrostatics.)  None of the three laws presented above are dynamical laws.  To get dynamics, we would also need a version of Newton's second law.  For rigid bodies, one can integrate the force density over the volume of the body to get a net force and use this to determine how the center of mass moves.  One would also need a law connecting the density of force on the body to its rotation.  For continua, the reaction to experienced forces will depend on the nature of the body experiencing the forces.  (We will return to that issue in section \ref{EMsection2}.)

Eliminating the gravitational field, you could alternatively have rigid bodies or continua interacting by gravitational action-at-a-distance---as in \eqref{lawofgravitation}.  All of this begins to illustrate the plethora of options that you have for understanding the laws and ontology of Newtonian gravity while staying within the time evolution paradigm.\footnote{There are also ways of formulating the laws that would fall outside the time evolution paradigm, as in the Lagrangian stationary action approach (\citealp{wharton2015}; \citealp[pg.\ 3--4, 35]{adlam2022}; \citealp[pg.\ 46--47]{chengoldstein}).}

\section{Electromagnetism, First Pass}\label{EMsection1}

To formulate the laws of electromagnetism, we must answer a central question about the theory's ontology:  Does the electromagnetic field really exist or is it merely a tool that can be used to calculate the forces between bits of charged matter?  One way to eliminate the electromagnetic field is to formulate electromagnetism as a theory of retarded action-at-a-distance\footnote{See \citet{lange, frischpietsch2016, absorbingthearrow}.} where the force on a given particle now is the result of interactions  with other particles at earlier times---the times when their world-lines intersect the past light-cone of the given particle now.  This version of electromagnetism breaks the time evolution paradigm because the interactions are not temporally local.  Another way to eliminate the electromagnetic field is to include half-retarded half-advanced action-at-a-distance, as in the Wheeler-Feynman version of the theory where the force on a given particle now is the result of interactions with other particles when and where they intersect either the past or future light-cones of the given particle.  Again, we have a theory that violates the time evolution paradigm because it is not temporally local.

Although the retarded action-at-a-distance and Wheeler-Feynman versions of electromagnetism are worthy of study, there are good reasons to prefer a version of electromagnetism that takes the electromagnetic field to be real.  Here are three:  First, a real electromagnetic field can ensure conservation of energy and momentum by itself possessing energy and momentum.\footnote{\citet[sec.\ 4.2]{lazarovici2018} gives a response to this concern defending the Wheeler-Feynman theory.}$^,$\footnote{\citet{lange} presents the argument for the reality of the electromagnetic field from conservation of energy and momentum, but finds it lacking in a relativistic context because energy and momentum are frame-dependent quantities.  He bases his argument for the field's reality on the fact that it possesses mass.  (Physicists classify the electromagnetic field as a massless field because the associated particle---the photon---is massless, but there is an important sense in which the field carries mass in addition to energy and momentum.  See \citealp[ch.\ 8]{lange}; \citealp[sec.\ 2]{gravitationalfield}.)}  Second, in general relativity the electromagnetic field acts as a source for gravitation. Third, in quantum electrodynamics the electromagnetic field is treated in a similar way to charged matter and thus it is natural (with the benefit of hindsight) to view both as real in classical electromagnetism (\citealp{fundamentalityoffields}; \citealp{absorbingthearrow}).\footnote{Proponents of the Wheeler-Feynman theory can of course pursue a revisionary approach to quantum field theory that eliminates the electromagnetic field and thus does not treat it in a similar way to charged matter.}

Accepting the electromagnetic field as real, one can still debate whether the field's behavior can always be traced back to past sources or whether the field has independent degrees of freedom.  Put more precisely, one can debate whether the electromagnetic field must obey the Sommerfeld radiation condition, restricting it to be the kind of electromagnetic field that you might introduce as a calculational tool within a retarded action-at-a-distance version of electromagnetism.  If you do impose such a restriction, the electric and magnetic fields at a given point in space and time can be determined by looking at the behavior of charged matter along the entire past light-cone of that point via Jefimenko's equations \citep[sec.\ 10.2.2]{griffiths}.  \citet[ch.\ 1]{jefimenko2000} criticizes Maxwell's equations because none fit the following mold:
\begin{quote}
``\dots equations depicting causal relations between physical phenomena must, in general, be equations where a present-time quantity (the effect) relates to one or more quantities (causes) that existed at some previous time.''
\citep[pg.\ 4]{jefimenko2000}
\end{quote}
Jefimenko takes Maxwell's equations to express relations between present-time quantities.  (We will see shortly that, by using future derivatives, two of Maxwell's equations can be understood as linking cause and effect with cause preceding effect.\footnote{This response to Jefimenko is similar to the response that \citet[pg.\ 853]{easwaran2014} gives to \citeauthor{field2003}'s \citeyearpar[pg.\ 438]{field2003} general claim that the differential equations that appear in laws of physics cannot be interpreted causally because they relate quantities at the same time.}) Unlike Maxwell's equations, Jefimenko's equations connect present fields to past sources and thus straightforwardly fit his mold.  However, that connection is temporally non-local and thus, although they may conform to our basic expectation that causes should come before effects, Jefimenko's equations do not fit the time evolution paradigm as stated in section \ref{TEPsection}.

The restricted version of electromagnetism that Jefimenko prefers forbids unsourced radiation.  It thus appears to yield a natural explanation as to why electromagnetic waves diverge---an explanation of the arrow of electromagnetic radiation.  That asymmetry can also be explained in an unrestricted version of electromagnetism, as it is improbable that any unsourced contribution to the electromagnetic field would result in coordinated converging waves \citep{north2003, absorbingthearrow}.  There is room to debate which version of electromagnetism gives a better explanation of this asymmetry, but the explanation offered by an unrestricted version has the advantage that the same kind of explanation can be given as to why entropy increases and why other waves diverge.  An unrestricted version of electromagnetism also fits well with quantum field theory, where the quantum electromagnetic field is not restricted to always have past sources (where, put another way, photons no more need past sources than electrons do).\footnote{See \citet[sec.\ 1.3]{wald2022}; \citet[sec.\ 3.3]{absorbingthearrow}.}

Having focused our attention on versions of electromagnetism that treat the electromagnetic field as real and do not impose a radiation condition requiring that field be traceable to past sources, there is still plenty of room to consider different versions of the theory.  \citet{maudlin2018} presents 10 distinct proposals for the laws and ontology.  It is not hard to find a version of the theory that fits the time evolution paradigm.  One way to do so is to start with the standard combination of Maxwell's equations and the Lorentz force law,\footnote{See, e.g., \citet[table 18-1]{feynman2}.} written here in Gaussian units:
\begin{align}
&\vec{\nabla}\cdot\vec{E}=4\pi \rho
\label{gauss}
\\
&\vec{\nabla}\times\vec{E}=-\frac{1}{c}\frac{\partial \vec{B}}{\partial t}
\label{faraday}
\\
&\vec{\nabla}\cdot\vec{B}=0
\label{gaussm}
\\
&\vec{\nabla}\times\vec{B}=\frac{4\pi}{c}\vec{J}+\frac{1}{c}\frac{\partial \vec{E}}{\partial t} 
\label{ampere}
\\
&\vec{F} = q \left(\vec{E} + \frac{1}{c}\vec{v} \times\vec{B} \right)
\ .
\label{lorentzforce}
\end{align}
If we understand the time derivatives in \eqref{faraday} and \eqref{ampere} to be future time derivatives, then those two equations can be dynamical laws for the electric and magnetic fields.  The other three laws would be non-dynamical laws.  To determine the future time evolution, the Lorentz force law \eqref{lorentzforce} would need to be accompanied by a dynamical law for particle motion, like Newton's second law \eqref{secondlaw} (modified to be made relativistic, as will be discussed in due course).

Note that here we are, for the moment, adopting an ontology of point particles for the charged matter and thus the charge and current densities in \eqref{gauss} and \eqref{ampere} will feature delta functions (as in \citealp[sec.\ 28]{landaulifshitzfields}).  The use of point particles will lead to trouble with self-interaction, as in section \ref{CMsection},  because the electromagnetic field will become infinitely strong at the location of any particle (just where the field's value matters for calculating forces).  Let us set these problems aside for now and return to them in section \ref{EMsection2}.

\section{The Relativistic Time Evolution Paradigm}\label{RTEPsection}

Although it is not difficult to find a version of electromagnetism that fits the time evolution paradigm, that paradigm may be too loose.  Electromagnetism is a relativistic theory and, as such, we can put tighter constraints on the form that the dynamical laws should take.  They should be spatiotemporally local,\footnote{Here I am talking about spatiotemporal locality in the special relativistic sense, not the broader sense in which there is no light-speed limit on influences and the only constraint is that there be spatiotemporal continuity (see \citealp[ch.\ 1]{lange}).} not merely temporally local.  Here is how \citet[pg.\ 60]{maudlin2007} puts the requirement of relativistic locality: ``the physical state at any point of space-time is determined or influenced only by events in its past light-cone.''\footnote{See also \citet[pg.\ 20 \& 30]{maudlin2007}.}  If the state at any point can only be influenced by events in its past light-cone, then the state at a particular space-time point can only influence events in the future light-cone of that point (for these are the events that it is in the past light-cone of).  Given Bell's theorem, quantum physics may force us to give up on this kind of locality (a point that we will return to in section \ref{QFTGRsection}).  But, it seems like this kind of locality should at least be achievable within classical electromagnetism.

Temporally local dynamical laws had to connect present and past neighborhood properties at a time to past-to-future neighborhood properties.  One might expect relativistic dynamical laws to connect space-time point properties and past light-cone neighborhood properties to past-to-future light-cone neighborhood properties.  Those terms can be defined as follows (modifying the definitions from section \ref{TEPsection}):
\begin{quote}
A \textbf{past light-cone neighborhood property} of an object at $(\vec{x},t)$ is not determined by the space-time point properties of the object at $(\vec{x},t)$ alone, but is determined once the space-time point properties of the object are specified within the past light-cone of $(\vec{x},t)$ over any arbitrarily small interval $[t-\Delta,t ]$.
\end{quote}
\begin{quote}
A \textbf{past-to-future light-cone neighborhood property} of an object at $(\vec{x},t)$ is not determined by the space-time point and past light-cone neighborhood properties of the object at $(\vec{x},t)$ alone, but is determined once the space-time point and past light-cone neighborhood properties of the object are specified within the future light-cone of $(\vec{x},t)$ over any arbitrarily small interval $[t,t+\Delta ]$.
\end{quote}
One can define future light-cone neighborhood property in parallel to past light-cone neighborhood property.  It follows from the definitions that future light-cone neighborhood properties are past-to-future light-cone neighborhood properties.

With these definitions in hand, let us say that a set of laws fits \textbf{The Relativistic Time Evolution Paradigm} if and only if the following three conditions are met (with changes from the time evolution paradigm in section \ref{TEPsection} marked in italics):
\begin{enumerate}
\item \textbf{\emph{Spatiotemporally} Local Dynamical Laws:}  A subset of the laws (the dynamical laws) can be applied at any \emph{point in space-time}, take as input only \emph{space-time point} properties or past \emph{light-cone} neighborhood properties, and give as output only past-to-future \emph{light-cone} neighborhood properties.  The dynamical laws may give either precise values for the past-to-future \emph{light-cone} neighborhood properties or probability distributions over such values.
\item \textbf{Non-Dynamical Laws:}  The laws that do not take the above dynamical form (the non-dynamical laws) express relations between \emph{space-time point} properties or past \emph{light-cone} neighborhood properties that hold at every \emph{space-time point}.
\item \textbf{\emph{Locally} Deterministic or Stochastic:} Once a law-abiding history over an arbitrarily small time interval \emph{within the past light-cone of some future spacetime point} has been fixed, the laws either uniquely fix a single future sequence of states \emph{within that light-cone} or the laws yield a precise probability distribution over future sequences of states \emph{within that light-cone}.
\end{enumerate}
Present properties (that make no reference to other times) have been replaced by space-time point properties (that make no reference to other times or places).  Past neighborhood properties have been replaced by past light-cone neighborhood properties (that are defined in terms of arbitrarily small subregions of the past light-cone).  Past-to-future properties have been replaced by past-to-future light-cone properties.  The third condition requires a local form of determinism or stochasticity.  What is going on at a time within any spherical region (and its arbitrarily-short past light-cone) should either fix, or give probabilities for, what occurs within the remainder of the  contracting light-cone that the spherical region is a slice of.\footnote{For an example as to how such local determinism can be proven, see the analysis of the source-free homogeneous wave equation in \citet[sec.\ 9.1]{strauss2008}.}

\citet{dorst} has criticized the dynamic production account of laws because it claims that earlier states produce later ones and this appears to require a preferred foliation---that is, a preferred way of carving space-time into simultaneity slices so that we can identify which slices are producing which.\footnote{\citet[pg.\ 46]{chengoldstein} give a quick version of this criticism as well.}  Looking at Bell's theorem in quantum physics, one might argue that we have independent reason to believe in a preferred foliation and be sanguine about its use here (\citealp[pg.\ 117]{maudlin2007}; \citealp[sec.\ 6]{maudlin2018}).  The relativistic time evolution paradigm provides an alternative way to respond to Dorst's worry.  We can view the dynamic laws as producing the future at each space-time point, with the inputs lying within the past-light-cone and the outputs concerning the future light-cone.  We can then collect these instances of production at-a-point to get production from one time slice to the next.  Different foliations will correspond to different ways of collecting these instances of production.

\begin{figure}[htb]
\center{\includegraphics[width=13 cm]{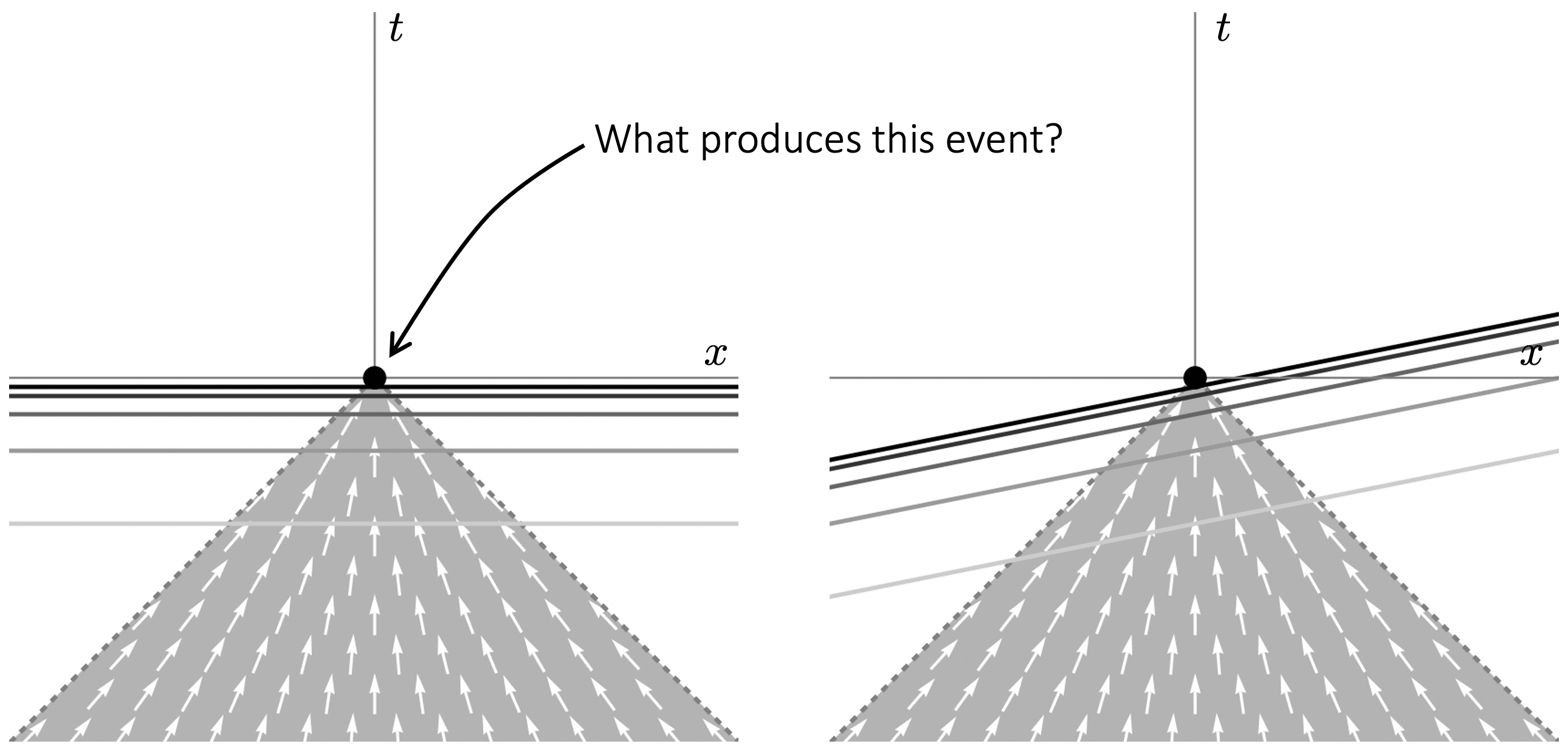}}
\caption{This figure shows schematically the chains of production connecting points in some event $e$'s past light-cone to the event.  It also shows two series of time slices approaching the event $e$ that can be described as producing the event.  The points in the event's past light-cone are carved up differently by the different series, but the underlying relations of production remain unchanged.}
  \label{cuts}
\end{figure}

I believe that Dorst would categorize this response as what he calls ``pluralistic production,'' because for every foliation it is true that the earlier states produce the later states.  Dorst rejects this response because if we ask what produces a given event $e$ at some particular space-time point, we get too many answers---overdetermination.  Can it really be that the previous time slices approaching the event $e$ on one foliation produce the event \emph{and} the previous time slices approaching the event on another foliation produce the event?  This may look like overdetermination, but because dynamic production is being understood here as fundamentally production at-a-point, it is really the space-time points that are shared between these two series that produce $e$.  The points in the past light-cone of $e$ produce $e$.  There are different ways to take a limit within this light-cone, but they capture the same chains of production (see figure \ref{cuts}).

\section{Electromagnetism, Second Pass}\label{EMsection2}

It requires a bit of work to fit electromagnetism into the relativistic time evolution paradigm.  The theory is normally understood as one where what happens at a given spacetime point is only influenced by its past light-cone and only influences its future light-cone, and thus normally understood to broadly fit the relativistic time evolution paradigm.  However, precisely fitting the theory into that paradigm, as articulated in section \ref{RTEPsection}, requires that the laws be spelled out with unusual care as to how the derivatives are understood so that, at every point in space-time, the laws link the past light-cone to the future light-cone.  The situation here is similar to the case of Newtonian gravity (section \ref{CMsection}), where the theory is normally understood as broadly fitting into the time evolution paradigm, but it takes some work to be careful about past and future derivatives so that the theory can be fit precisely into the time evolution paradigm from section \ref{TEPsection}.  Electromagnetism is a more complex theory than Newtonian gravity, and the relativistic time evolution paradigm is a stricter standard than the time evolution paradigm, so more work is needed here than there.

Below I present one way to fit electromagnetism into the time evolution paradigm, without claiming that it is the best and final way.  We need laws governing the evolution of the electromagnetic field, the forces on charged matter, and the reaction of matter to forces.  Let us begin with the field.

\subsection{The evolution of the electromagnetic field}

We can use the scalar potential $\phi$ and the vector potential $\vec{A}$ to specify the state of the electromagnetic field, where these potentials are related to $\vec{E}$ and $\vec{B}$ by
\begin{align}
\vec{E}&=-\vec{\nabla}\phi-\frac{1}{c}\frac{\partial \vec{A}}{\partial t}
\nonumber
\\
\vec{B}&=\vec{\nabla} \times \vec{A}
\ .
\label{fieldsfrompotentials}
\end{align}
These relations ensure that two of Maxwell's equations are satisfied automatically, \eqref{faraday} and \eqref{gaussm}.  Let us adopt the Lorenz gauge condition as a way of partially fixing the gauge freedom in the potentials,\footnote{Adopting the Coulomb gauge would lead to very different laws for the potentials, laws that appear to involve instantaneous action-at-a-distance and to rely on a preferred simultaneity slicing \citep{maudlin2018}.  Our choice to use vector and scalar potentials in the Lorenz gauge in this section is not merely a choice of convention.  It is part of a substantive proposal about the ontology and laws of electromagnetism attempting to fit electromagnetism into the relativistic time evolution paradigm.}
\begin{equation}
\vec{\nabla}\cdot\vec{A}+\frac{1}{c}\frac{\partial \phi}{\partial t}=0
\ .
\end{equation}
The remaining two of Maxwell's equations yield wave equations for $\phi$ and $\vec{A}$,
\begin{align}
\left(\nabla^2-\frac{1}{c^2}\frac{\partial^2}{\partial t^2}\right)\phi&=-4\pi\rho
\label{phiwave}
\\ 
\left(\nabla^2-\frac{1}{c^2}\frac{\partial^2}{\partial t^2}\right)\vec{A}&=-\frac{4\pi}{c}\vec{J}
\ ,
\label{awave}
\end{align}
which can be written more compactly using the d'Alembertian, $\Box=\nabla^2-\frac{1}{c^2}\frac{\partial^2}{\partial t^2}$, as
\begin{align}
\Box\phi&=-4\pi\rho
\\ 
\Box\vec{A}&=-\frac{4\pi}{c}\vec{J}
\ .
\end{align}

A first try at formulating these wave equations as laws of time evolution for the potentials, modeled on Easwaran's version of Newton's second law \eqref{secondlaw}, would be to interpret the second derivative with respect to time as a future derivative of a past derivative (just as acceleration wast taken to be the future derivative of the past derivative of position),
\begin{align}
\left(\frac{\partial}{\partial t}\right)^f\left(\frac{\partial}{\partial t}\right)^p\phi&=4\pi c^2\rho-c^2\nabla^2\phi
\\
\left(\frac{\partial}{\partial t}\right)^f\left(\frac{\partial}{\partial t}\right)^p\vec{A}&=4\pi c\vec{J}-c^2\nabla^2\vec{A}
\ .
\end{align}
These equations would yield temporally local but not spatiotemporally local dynamics.  The $\nabla^2$ operators yield properties of the potentials that are determined by considering arbitrarily small spatial neighborhoods at the moment in question.  That is forbidden by the relativistic time evolution paradigm.  These are not space-time point properties or past light-cone neighborhood properties.

Although this first attempt fails, we should be optimistic that the wave equations in \eqref{phiwave} and \eqref{awave} can be interpreted as giving spatiotemporally local dynamics.  When the homogenous (source-free) wave equation, $\Box u = 0$, is discussed in textbooks on partial differential equations, it is standard practice to prove a ``causality theorem'' showing that a given spacetime point cannot influence anything outside its future light-cone and cannot be influenced by anything outside its past light-cone.\footnote{For discussion of causality theorems for one and three-dimensional wave equations, see \citet[ch.\ 8]{zachmanoglou1976}; \citet[ch.\ 5]{folland1995}; \citet[sec.\ 2.4.3]{evans1998}; \citet[sec.\ 9.1]{strauss2008}.  For discussion of causality theorems in electromagnetism (making use of the Lorenz gauge), see \citet[sec.\ 10.2]{wald1984}; \citet[sec.\ 5.4]{wald2022}.\label{causalitytheorems}}  To be more precise, you can prove that if two solutions agree on the values of the function $u$ and its time derivatives $\frac{\partial u}{\partial t}$ at a given time within the spherical spatial region bounded by the past light-cone of some future spacetime point $(\vec{x},t)$, the two solutions will agree on the value of $u$ at $(\vec{x},t)$.  Solutions to the inhomogeneous (sourced) wave equation, $\Box u = f$, will also satisfy such a causality theorem for a fixed source $f$ because any solution can be divided into a free part $u_{in}$ (that obeys the homogenous wave equation and can be interpreted as describing incoming waves that are not attributable to the source function $f$) and a retarded part $u_{ret}$ (where the value of $u_{ret}$ at any space-time point is fixed by $f$ along the past light-cone of that point).  Given these causality results, it seems like there should be a way to formulate inhomogeneous wave equations, like the wave equations for the potentials \eqref{phiwave} and \eqref{awave}, that makes the causal structure manifest and satisfies the conditions that we laid out earlier for spatiotemporally local dynamics linking past light-cone to future light-cone.  As we will see, this can be done straightforwardly in one-dimensional space but is surprisingly difficult in three-dimensional space.

Let us begin by considering the inhomogeneous wave equation in one-dimensional space,
\begin{equation}
\left(\frac{\partial^2}{\partial x^2}-\frac{1}{c^2}\frac{\partial^2}{\partial t^2}\right)u=f
\ ,
\label{1Dwave}
\end{equation}
with $f$ treated as a fixed source function defined across space and time.  D'Alembert's method for solving the homogeneous wave equation introduces new variables\footnote{These are one-dimensional versions of Dirac's light-cone, or light-front, coordinates \citep{dirac1949}.}
\begin{align}
\eta &= x+ct
\nonumber
\\
\xi&=x-ct
\ .
\label{1Dvars}
\end{align}
We can use these variables to rewrite \eqref{1Dwave} as
\begin{equation}
4 \frac{\partial}{\partial \eta} \frac{\partial}{\partial \xi}u=f
\ ,
\label{1DLCderivatives}
\end{equation}
where the derivatives are evaluated along the two perpendicular edges of the light-cone.  If you were to take both derivatives to be future derivatives, then the output (the left-hand side of the equation) would be a future light-cone neighborhood property (and thus a past-to-future light-cone neighborhood property) and the input (the right-hand side) would be a space-time point property (assuming that $f$ gives space-time point properties).  However, this interpretation of \eqref{1DLCderivatives} would not capture the causal structure accurately. The source function $f$ alone does not determine how $u$ will change towards the future.  At a moment, you need $f$, $u$, and the rate at which $u$ is changing to fix the future evolution.

\begin{figure}[htb]
\center{\includegraphics[width=14 cm]{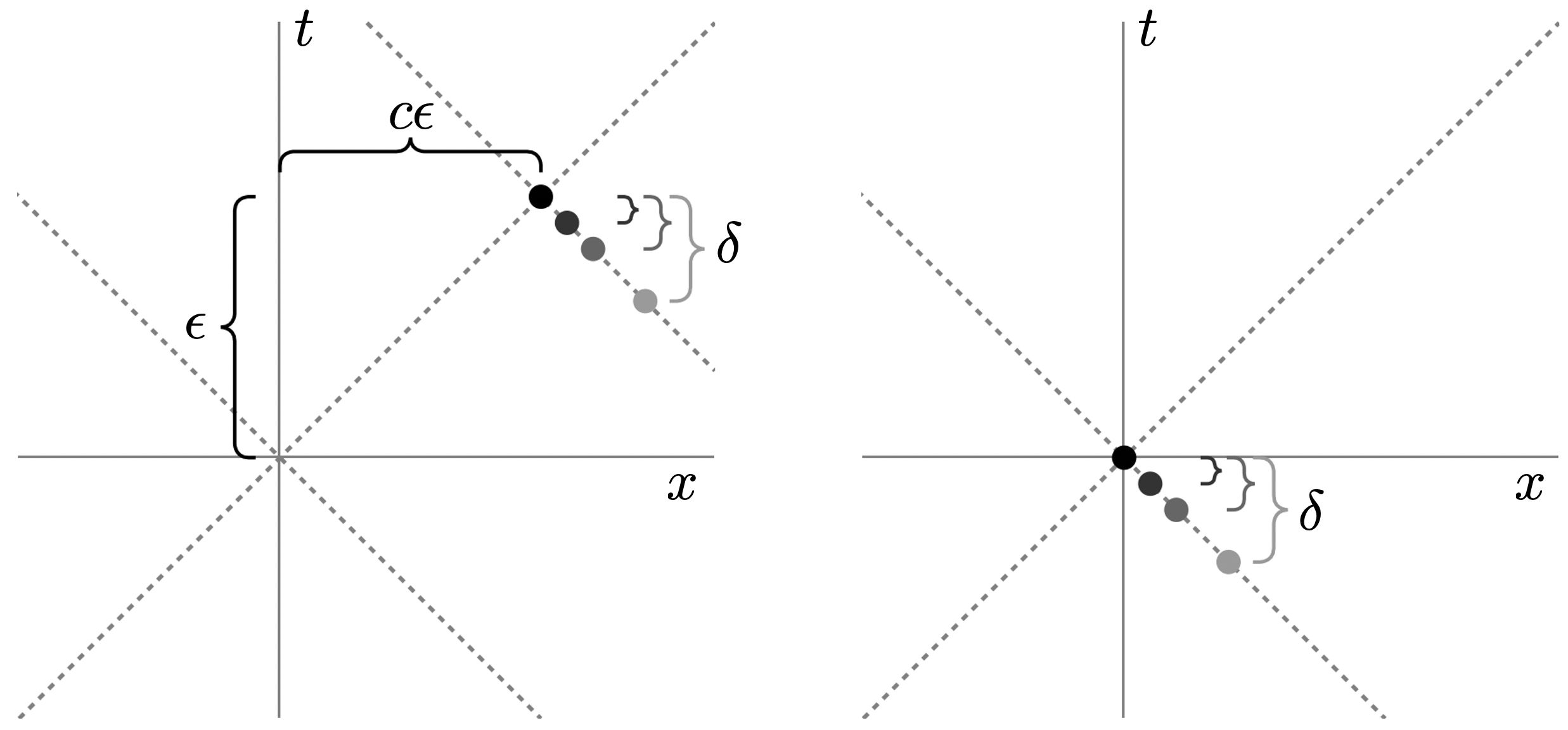}}
\caption{If we rewrite the one-dimensional wave equation using D'Alembert's variables as in \eqref{1DLCFderivatives} and \eqref{1DLCFderivatives2}, then one must compare (a) how the values of the function $u$ change as you approach a spacetime point $\epsilon$ up the right edge of the future light-cone (the black, highest dot in the left image), coming progressively closer to that point along the right edge of the past light-cone (the darkening gray dots in the left image) to (b) the way the values of $u$ change as you approach the spacetime point of interest (the black dot in the right image), coming progressively closer to that point along the right edge of the past light-cone (the darkening gray dots in the right image).  The left image depicts the first two terms in the numerator of \eqref{1DLCFderivatives2} and the right image depicts the last two terms.}
  \label{limits}
\end{figure}

Taking the first derivative in \eqref{1DLCderivatives} to be a future derivative and the second to be a past derivative, the wave equation becomes
\begin{equation}
4 \left(\frac{\partial}{\partial \eta}\right)^f \left(\frac{\partial}{\partial \xi}\right)^p u=f
\ .
\label{1DLCFderivatives}
\end{equation}
The past derivative can be written as a limit,
\begin{equation}
4 \left(\frac{\partial}{\partial \eta}\right)^f  \left(\lim_{\delta \to 0}\frac{u(x+c\delta, t-\delta)-u(x, t)}{2 c\delta}\right)=f
\ ,
\end{equation}
with the factor of 2 in the denominator arising from the fact that a step of $c\delta$ for $x$ and $-\delta$ for $t$ is a step of $2c\delta$ for $\xi$.  Writing the remaining derivative as a limit yields,
\begin{equation}
\lim_{\epsilon \to 0} \left(\lim_{\delta \to 0}\frac{u(x+c\epsilon+c\delta, t+\epsilon-\delta)-u(x+c\epsilon, t+\epsilon)-u(x+c\delta, t-\delta)+u(x, t)}{c^2\epsilon\delta}\right)=f
\ .
\label{1DLCFderivatives2}
\end{equation}
This expression asks you to go up one edge of the future light-cone shorter and shorter distances, each time taking a limit toward that edge of the light-cone approaching from the past along the perpendicular to the future light-cone (see figure \ref{limits}).  This counts as a spatiotemporally local dynamical law because the left-hand side gives a past-to-future light-cone neighborhood property as output and the right-hand side gives what might be a space-time point property or a past light-cone neighborhood property, depending on the source function $f$ (as we will see below).  Note that the output is not a future light-cone neighborhood property because at each point along the future light-cone that you consider for the derivative with respect to $\eta$, you must look outside the light-cone to evaluate $\left(\frac{\partial}{\partial \xi}\right)^p u$ (as is shown in figure \ref{limits}).\footnote{We can now follow up on a loose thread from footnote \ref{openended1}: Even if we were to follow \citet{easwaran2014} and use open-ended derivatives instead of closed-ended derivatives, the output of \eqref{1DLCFderivatives} would not be a future light-cone neighborhood property because the past derivative with respect to $\xi$ takes you outside the future light-cone.\label{openended2}}

Looking back at \eqref{1DLCFderivatives}, you could just as well swap the order of the derivatives and go up the other edge of the future light-cone.  To get a more symmetric law and avoid privileging either side of the future light-cone, you could average the two orderings:
\begin{equation}
2 \left(\frac{\partial}{\partial \eta}\right)^f \left(\frac{\partial}{\partial \xi}\right)^p u+2 \left(\frac{\partial}{\partial \xi}\right)^f \left(\frac{\partial}{\partial \eta}\right)^p u=f
\ .
\label{1DLCFderivatives3}
\end{equation}
Here we have arrived at an appealing formulation of the one-dimensional wave equation as a spatiotemporally local dynamical law.

In three-dimensional space, the inhomogeneous wave equation becomes
\begin{equation}
\left(\nabla^2-\frac{1}{c^2}\frac{\partial^2}{\partial t^2}\right)u=f
\ ,
\label{3Dwave}
\end{equation}
where we will  again treat $f$ as a fixed source function defined across space and time.  The wave equations for the potentials, \eqref{phiwave} and \eqref{awave}, fit this general form.  Writing this wave equation as a spatiotemporally local dynamical law is not as straightforward as you might expect, but here is one way to do it.  Let us focus our attention on the evolution at the origin, noting that any point in space can be treated as the origin.  At the origin,\footnote{Away from the origin, the Laplacian cannot be put in this form and must include derivatives with respect to $\theta$ and $\phi$ in addition to $r$ (as is evident in the standard expression for the Laplacian in spherical coordinates).} the Laplacian $\nabla^2=\frac{\partial^2}{\partial x^2}+\frac{\partial^2}{\partial y^2}+\frac{\partial^2}{\partial z^2}$ can be rewritten in spherical coordinates as an integral over contributions from second derivatives with respect to the radial coordinate $r$ for different choices of $\theta$ (the azimuthal angle) and $\phi$ (the polar angle)---seeing how $u$ changes as you look out in different directions from the origin:\footnote{I have not seen \eqref{laplacianspherical} presented elsewhere. To derive \eqref{laplacianspherical}, plug the expression for the radial derivative given a particular choice of $\theta$ and $\phi$,
\begin{equation}
\frac{\partial}{\partial r}=\cos\theta\sin\phi \frac{\partial}{\partial x}+\sin\theta\sin\phi \frac{\partial}{\partial y}+\cos\phi \frac{\partial}{\partial z}
\ ,
\label{rderiv}
\end{equation}
into the integral over these angles on the right-hand side of \eqref{laplacianspherical}.  When you perform the integrals over $\theta$ and $\phi$, the cross terms drop out and you are left with the Laplacian (at the origin):
  \begin{align}
3 \int \frac{d\theta d\phi}{4\pi}\sin\phi \; \frac{\partial^2}{\partial r^2}u&=3 \int \frac{d\theta d\phi}{4\pi}\bigg(\cos^2\theta\sin^3\phi \frac{\partial^2}{\partial x^2}+\sin^2\theta\sin^3\phi \frac{\partial^2}{\partial y^2}+\cos^2\phi\sin\phi \frac{\partial^2}{\partial z^2}
\nonumber
\\
&\quad+2\cos\theta\sin\theta\sin^3\phi \frac{\partial}{\partial x}\frac{\partial}{\partial y}+2\sin\theta\sin^2\phi\cos\phi \frac{\partial}{\partial y}\frac{\partial}{\partial z}+2\cos\theta\sin^2\phi\cos\phi \frac{\partial}{\partial x}\frac{\partial}{\partial z}\bigg)u
\nonumber
\\
&=\left(\frac{\partial^2}{\partial x^2}+\frac{\partial^2}{\partial y^2}+\frac{\partial^2}{\partial z^2}\right)u=\nabla^2 u
\ .
 \end{align}}
\begin{equation}
\nabla^2 u = 3 \int \frac{d\theta d\phi}{4\pi}\sin\phi \; \frac{\partial^2}{\partial r^2}u
\ ,
\label{laplacianspherical}
\end{equation}
where $\int \frac{d\theta d\phi}{4\pi}\sin\phi$, or $\int \frac{d\Omega}{4\pi}$, integrates over the solid angle $\Omega$ and divides by the total solid angle.  If we include a trivial integral over the solid angle for the time derivative term, the wave equation \eqref{3Dwave} for $u$ at the origin becomes
  \begin{align}
\int \frac{d\theta d\phi}{4\pi}\sin\phi \left(3 \frac{\partial^2}{\partial r^2}-\frac{1}{c^2}\frac{\partial^2}{\partial t^2}\right)u&=f
\ ,
\label{3Dwavespherical}
 \end{align}
 which looks more like the one-dimensional wave equation \eqref{1Dwave}.
 
\begin{figure}[htb]
\center{\includegraphics[width=10 cm]{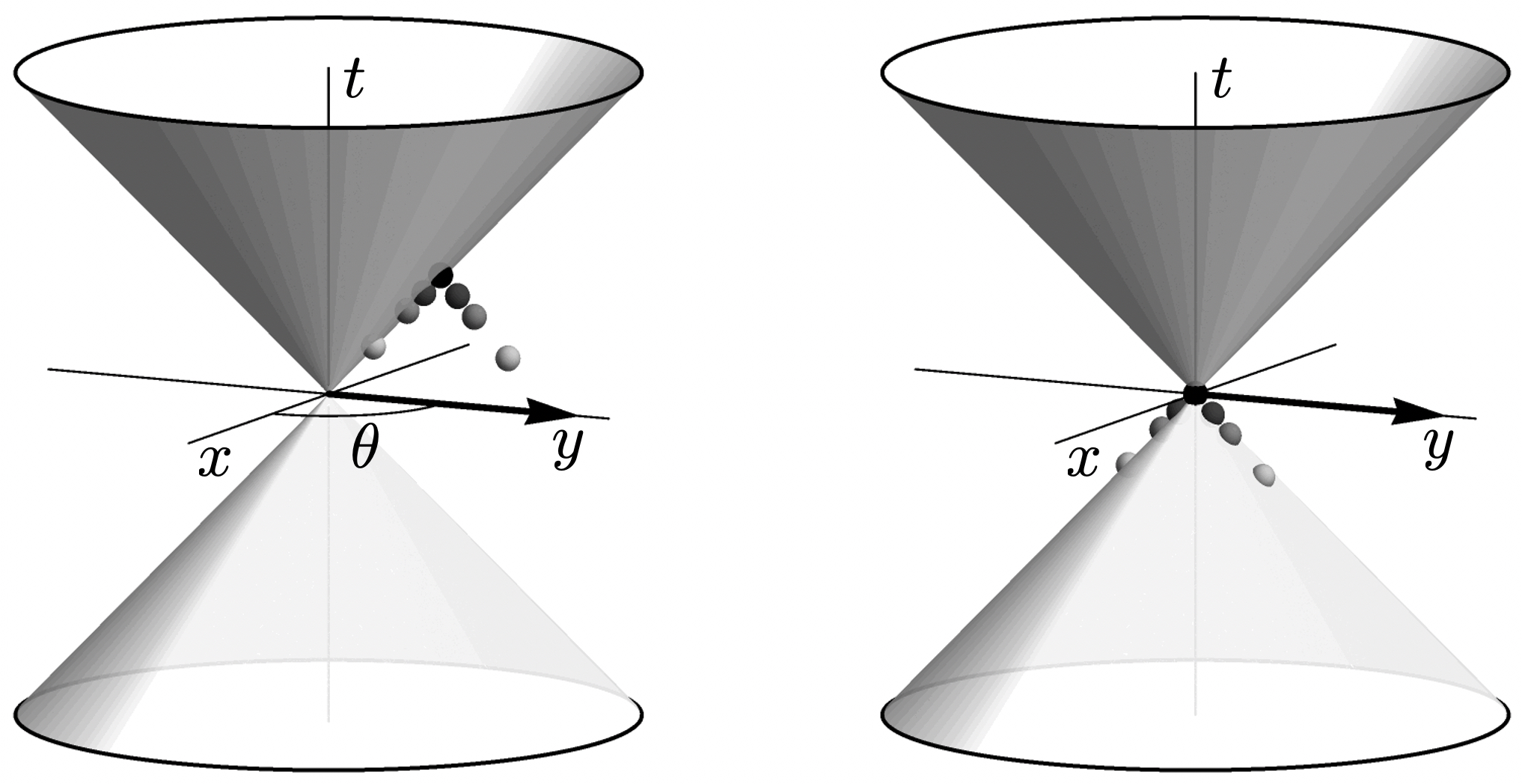}}
\caption{The spatiotemporally local form of the three-dimensional wave equation \eqref{3Dwavelaw} asks us to compare, for each direction picked out by a set of angles $\theta$ and $\phi$, (a) how the values of the function $u$ are changing as you approach a point displaced from the point of interest by an arbitrarily amount along that direction into the future light-cone from that direction along the past light-cone and the opposite direction (left image) to (b) how the values of $u$ are changing as you approach the point of interest from that direction along the past light-cone and the opposite direction (right image).  The two images only show the comparison for a single direction, $\theta=\frac{\pi}{2}$ and $\phi=\frac{\pi}{2}$ (the $y$ direction), and depict only the $x$, $y$, and $t$ dimensions.  The darkening dots show different points where $u$ would be evaluated in calculating the light-cone derivatives for this direction.}
  \label{limits3D}
\end{figure}

We can now introduce new variables going out along the future and past light-cones,
  \begin{align}
\eta &= r+ct
\nonumber
\\
\xi &=r-ct
\ ,
\label{3Dvars}
\end{align}
similar to D'Alembert's variables \eqref{1Dvars} from the one-dimensional case.  Using these variables, the wave equation at the origin \eqref{3Dwavespherical} is
  \begin{align}
\int \frac{d\Omega}{4\pi} \left(2\frac{\partial^2}{\partial \eta^2}+2\frac{\partial^2}{\partial \xi^2}+8\frac{\partial^2}{\partial \eta \partial \xi}\right)u&=f
\ .
\label{3Dwavespherical2}
 \end{align}
 In the integral, for a given $\theta$ and $\phi$, the $\frac{\partial^2}{\partial \eta^2}$ term asks you to take a second derivative along the future light-cone (in the direction picked out by $\theta$ and $\phi$) and the $\frac{\partial^2}{\partial \xi^2}$ term asks you to take a second derivative along the past light-cone (in that direction).  This past light-cone derivative should give the same result as taking the forward light-cone derivative in the opposite direction, $-\hat{r}$.  Integrating over all angles,\footnote{Here is a more formal proof of \eqref{etaxiidentity}.  Expanded in terms of $r$ and $t$ derivatives, the difference between the left and right-hand sides of \eqref{etaxiidentity} is proportional to
 \begin{equation}
 \int \frac{d\Omega}{4\pi} \left(\frac{\partial}{\partial r}\frac{\partial}{\partial t}\right)=0
 \ .
 \end{equation}
The fact that this term vanishes can be seen by noting that radial derivatives in opposite directions will cancel, or, by using the expansion of $\frac{\partial}{\partial r}$ in \eqref{rderiv}.
 }
   \begin{align}
\int \frac{d\Omega}{4\pi} \left(\frac{\partial^2}{\partial \eta^2}\right)=\int  \frac{d\Omega}{4\pi} \left(\frac{\partial^2}{\partial \xi^2}\right)
\ .
\label{etaxiidentity}
 \end{align}
 We can thus combine the first two terms in \eqref{3Dwavespherical2} to simplify the equation, yielding
   \begin{align}
4\int\frac{d\Omega}{4\pi} \left(\frac{\partial^2}{\partial \eta^2}+2\frac{\partial^2}{\partial \eta \partial \xi}\right)u&=f
\ .
 \end{align}
Regrouping gives
    \begin{align}
4\int \frac{d\Omega}{4\pi} \left(\frac{\partial}{\partial \eta}\left[\frac{\partial}{\partial \eta}+2\frac{\partial}{\partial \xi}\right]\right)u&=f
\ .
 \end{align}
Let us now distinguish past and future derivatives as in \eqref{1DLCFderivatives} and \eqref{1DLCFderivatives3},
\begin{align}
4\int \frac{d\Omega}{4\pi} \left(\left(\frac{\partial}{\partial \eta}\right)^f\left[\left(\frac{\partial}{\partial \eta}\right)^p+2\left(\frac{\partial}{\partial \xi}\right)^p\right]\right)u&=f
\ .
\label{3Dwavelaw}
 \end{align}
For given angles $\theta$ and $\phi$, the integrand asks to consider how a sum of two past light-cone derivatives (one in the $\hat{r}$ direction and one in the $-\hat{r}$ direction) changes as you move along the future light-cone in the $\hat{r}$ direction (see figure \ref{limits3D}).  With this form of the wave equation, we have arrived at a spatiotemporally local dynamical law where the left-hand side is a past-to-future light-cone neighborhood property and the source function on the right-hand side is either a space-time point property or a past light-cone property (as we are about to see).\footnote{One should not read the speed of causal influence directly from the ability to put the wave equation in a form like \eqref{3Dwavelaw}.  If you started with a wave equation like \eqref{3Dwave} with the speed of light of light doubled ($c \rightarrow 2 c$), local determinism would not hold (relative to ordinary light-cones) but you could follow parallel reasoning to the derivation above to get an expression for the wave equation that looks similar to \eqref{3Dwavelaw} and uses the same coordinates $\eta$ and $\xi$ from \eqref{3Dvars}.}
   
The wave equation for the scalar potential \eqref{phiwave} fits the form of a three-dimensional wave equation \eqref{3Dwave} and thus we can formulate it (at the origin) as a spatiotemporally local law like \eqref{3Dwavelaw},
     \begin{align}
\int \frac{d\Omega}{4\pi} \left(\left(\frac{\partial}{\partial \eta}\right)^f\left[\left(\frac{\partial}{\partial \eta}\right)^p+2\left(\frac{\partial}{\partial \xi}\right)^p\right]\right)\phi&=-\pi\rho
\ ,
\label{philaw}
 \end{align}
where the $\phi$ that appears here is the scalar potential, not to be confused with the azimuthal angle.  Here the density of charge $\rho$ acts as the source term.  Sometimes $\rho$ is thought of as a coarse-grained blurring of an underlying spiky charge distribution of distinct bodies.  Because it is a kind of average of the charge over a small region, you might worry that it involves looking outside of the past light-cone---and thus that \eqref{philaw} may fail to be a spatiotemporally local law.  Let us think of $\rho$ as fundamental, either including delta functions (if matter is modeled as point charges) or smoothly varying (if matter is modeled using charge distributions).  Then, we can take the value of $\rho$ at a point to be a space-time point property and \eqref{philaw} fits the mold of a spatiotemporally local dynamical law.
 
The wave equation for the vector potential \eqref{awave} is three separate three-dimensional wave equations of the earlier form \eqref{3Dwave}, one for each component of the potential.  We can formulate it (at the origin) as 
\begin{align}
\int \frac{d\Omega}{4\pi} \left(\left(\frac{\partial}{\partial \eta}\right)^f\left[\left(\frac{\partial}{\partial \eta}\right)^p+2\left(\frac{\partial}{\partial \xi}\right)^p\right]\right)\vec{A}&=-\frac{\pi}{c}\vec{J}
\ .
\label{alaw}
 \end{align}
For point charges, current density $\vec{J}$ that sources the vector potential can be formed by combining contributions of the form $\rho\vec{v}$ associated with each point charge, where $\rho$ is a delta function centered on the charge's location and $\vec{v}$ is the charge's velocity \citep[sec.\ 28]{landaulifshitzfields}.  As before, the velocity can be understood as a past derivative of the body's location and thus depends on an arbitrarily short segment of its past trajectory.  If the body has never moved faster than the speed of light, then this trajectory will lie within the past light-cone and the velocity will be a past light-cone neighborhood property.  The current density will thus also be a past light-cone neighborhood property.

What if charged matter is modeled instead by a charge distribution?  In that case, the current density at a moment cannot be read off from the present charge distribution and its recent history.  \citet[pg.\ 8]{maudlin2018} gives the example of a uniform sphere that maintains a uniform charge density.  This is compatible with there being no current density or with their being a current density corresponding to a rotation of the charge about some axis.  Thus, unlike the velocity of a point charge, the current density of a continuum cannot be reduced to the historic behavior of charge.  One option would be to treat the charge distribution and current density as distinct properties of the instantaneous state.  The current density then becomes a space-time point property, not a past light-cone neighborhood property.  Again, \eqref{alaw} counts as a spatiotemporally local dynamical law.

\subsection{The forces on charged matter}\label{FCMsection}

We have been attempting to fit electromagnetism into the relativistic time evolution paradigm and thus far we have focused on the laws governing the evolution of the electromagnetic field.  To complete the task, we must also analyze the forces on charged matter and the reaction of matter to these forces.  First, let us consider the Lorentz force law for point charges \eqref{lorentzforce}.  Using the potentials, we can rewrite this law as\footnote{See \citet[sec.\ 17]{landaulifshitzfields}; \citet[pg.\ 1365]{semontaylor}; \citet[pg.\ 442]{griffiths}.}
\begin{align}
\vec{F} &= -q\vec{\nabla}\phi-\frac{q}{c}\frac{\partial \vec{A}}{\partial t}+\frac{q}{c}\vec{v} \times \left(\vec{\nabla} \times \vec{A}\right)
\nonumber
\\
&= -q\vec{\nabla}\phi-\frac{q}{c}\frac{\partial \vec{A}}{\partial t}+\frac{q}{c}\left(\vec{\nabla}\left(\vec{v}\cdot\vec{A}\right) - \left(\vec{v}\cdot\vec{\nabla}\right)\vec{A}\right)
\nonumber
\\
&= - q\vec{\nabla}\left(\phi-\frac{1}{c}\vec{v}\cdot\vec{A}\right)-\frac{q}{c}\frac{D \vec{A}}{D t}
\ .
\label{lorentzforceexpansion}
\end{align}
In the last line, $\frac{D}{D t}=\frac{\partial}{\partial t} + (\vec{v}\cdot\vec{\nabla})$ is the convective derivative, a derivative taken along the path of the charge.  If we take this to be a past derivative and assume that the particle has not moved faster than the speed of light, then the final term is a past light-cone neighborhood property.  The remainder of the expression is a gradient that appears to require looking outside the past light-cone, and thus to cause trouble if we want the force to be a past light-cone neighborhood property so that the dynamics are spatiotemporally local.  There is a way to address this problem, though it might feel like a cheap trick.  In analogy with \eqref{1Dvars}, we can introduce coordinates $\eta_i = x_i+ct$ and $\xi_i=x_i-ct$ to replace the gradient and write the $x_i$-th component of the force as
\begin{align}
F_i &= - q\left[\frac{\partial}{\partial \eta_i}+\frac{\partial}{\partial \xi_i}\right]\left(\phi-\frac{1}{c}\vec{v}\cdot\vec{A}\right)-\frac{q}{c}\frac{D A_i}{D t}
\ ,
\end{align}
where $i$ is an index on the spatial dimensions and should not be confused with the earlier use of $i$ as an index on the charges.  Making all of the derivatives past derivatives gives the force law
\begin{align}
F_i &= - q\left[\left(\frac{\partial}{\partial \eta_i}\right)^p+\left(\frac{\partial}{\partial \xi_i}\right)^p\,\right]\left(\phi-\frac{1}{c}\vec{v}^{\,p}\cdot\vec{A}\right)-\frac{q}{c}\left(\frac{D}{D t}\right)^p A_i
\ .
\label{LCLorentz}
\end{align}
To clarify that the derivatives do not act on $\vec{v}^{\,p}$, this can be rewritten as
\begin{align}
F_i &= - q\left[\left(\frac{\partial}{\partial \eta_i}\right)^p+\left(\frac{\partial}{\partial \xi_i}\right)^p\,\right]\phi+\frac{q}{c}\vec{v}^{\,p}\cdot\left(\left[\left(\frac{\partial}{\partial \eta_i}\right)^p+\left(\frac{\partial}{\partial \xi_i}\right)^p\,\right]\vec{A}\right)-\frac{q}{c}\left(\frac{D}{D t}\right)^p A_i
\ .
\label{LCLorentz2}
\end{align}
In words: The force from the electromagnetic field in a given direction depends on (i) the way the scalar potential is changing forward and backward in that direction along the past light-cone, (ii) the dot product of the charge's velocity with the way the vector potential is changing forward and backward in that direction along the past light-cone, and (iii) the time derivative of the vector potential along the past trajectory of the charge.  Here we see that \eqref{LCLorentz2} has the form of a non-dynamical law relating past light-cone neighborhood properties.

Modeling matter as a continuous charge distribution, the above Lorentz force law for point charges \eqref{LCLorentz2} becomes an equation for the force density $f_i$ in the $x_i$-th direction,
\begin{align}
f_i &= - q\left[\left(\frac{\partial}{\partial \eta_i}\right)^p+\left(\frac{\partial}{\partial \xi_i}\right)^p\,\right]\phi+\frac{q}{c}\vec{v}\cdot\left(\left[\left(\frac{\partial}{\partial \eta_i}\right)^p+\left(\frac{\partial}{\partial \xi_i}\right)^p\,\right]\vec{A}\right)-\frac{q}{c}\left(\frac{D}{D t}\right)^p A_i
\ ,
\label{LCLorentzdensity}
\end{align}
where the velocity that appears here is a function of space and time.

\subsection{The reaction of matter to forces}

As the final piece of the puzzle, let us now discuss the way that charges respond to forces.  That is not always considered part of electromagnetism proper and does not appear in \eqref{gauss}--\eqref{lorentzforce}, but it would be needed to arrive at a complete theory that has a hope of fitting the relativistic time evolution paradigm.  For point charges, we might use Newton's second law \eqref{secondlaw}, $\vec{F}=m\vec{a}^{\,pf}$, as a non-relativistic approximation to the reaction of charges to forces.  The relativistic law would be
\begin{equation}
\vec{F}=\left(\frac{d}{dt}\right)^f \vec{p}^{\,p}
\ ,
\label{secondlawp}
\end{equation}
using a relativistic expression for the momentum,
\begin{equation}
\vec{p}^{\,p}=\frac{m\vec{v}^{\,p}}{\sqrt{1-\frac{|\vec{v}^{p}|^2}{c^2}}}
\ .
\label{relmomentum}
\end{equation}
Because the particles do not move faster than $c$, the momentum $\vec{p}^{\,p}$ will be a past light-cone neighborhood property and the right-hand side of \eqref{secondlawp}, $\left(\frac{d}{dt}\right)^f \vec{p}^{\,p}$, will be a past-to-future light-cone neighborhood property.  If we assume that we are dealing with purely electromagnetic forces then, as was discussed in section \ref{FCMsection}, the Lorentz force \eqref{lorentzforceexpansion} that would appear on the right-hand side of \eqref{secondlawp} is a past light-cone neighborhood property.  Thus, the reaction law \eqref{secondlawp} has the correct form to be a spatiotemporally local dynamical law.  As in section \ref{CMsection}, you could combine the non-dynamical Lorentz force law \eqref{LCLorentz2} and the dynamical reaction law \eqref{secondlawp} into a single dynamical law (assuming that the only forces at play are electromagnetic forces).  Making that move, the laws of electromagnetism would all be dynamical laws and there would be no need for any non-dynamical laws.

If the matter is modeled using continuous distributions of charge instead of point charges, the reaction of matter to forces is more complicated and will depend on the nature of the matter.  You would need to couple some theory of the matter with the above laws of electromagnetism to get a complete theory that could fit the relativistic time evolution paradigm.  A neatly relativistic way to do this would be to model the matter using a classical Klein-Gordon or Dirac field in a Maxwell-Klein-Gordon or Maxwell-Dirac classical field theory.  Then, one would need to show that the wave equations for these fields can be formulated in ways that fit the relativistic time evolution paradigm, as we have seen that the wave equations of electromagnetism can---\eqref{philaw} and \eqref{alaw}.  For the Klein-Gordon field, that is straightforward as we can simply carry over the techniques that were used for the electromagnetic field.\footnote{You can prove the same kind of causality theorem for the Klein-Gordon equation as the one we discussed for the wave equation near the beginning of this section (\citealp[sec.\ 10.1]{wald1984}; \citealp[pg.\ 234, problem 8]{strauss2008}).}  For the Dirac field, it is not so clear how to formulate the wave equation as a spatiotemporally local dynamical law.  Note that with either the Klein-Gordon or Dirac equation in place, the Lorentz force law \eqref{LCLorentzdensity} is not needed as a fundamental non-dynamical law (though it could still be regarded as a non-fundamental law giving the density of force exerted by the electromagnetic field on the Klein-Gordon or Dirac field\footnote{The idea that fields can experience forces is discussed in \citet{forcesonfields, spinmeasurement}.}).

\subsection{The whole package}

Putting it all together, we have arrived at a way of formulating electromagnetism that seems like it could fit the relativistic time evolution paradigm.  For point charges, we can take the laws to be given by the wave equations for the potentials, \eqref{philaw} and \eqref{alaw}, the Lorentz force law, \eqref{LCLorentz2}, and the relativistic version of Newton's second law \eqref{secondlawp}---or, \eqref{LCLorentz2} and \eqref{secondlawp} could be combined.  However, there is danger on the horizon for that kind of theory.  We could run into problems of self-interaction like those faced by Newtonian gravity with point masses (discussed in section \ref{CMsection}).

Just as the gravitational field becomes infinite as you approach a point mass and ill-defined at its location, the electric field becomes infinite as you approach a point charge and ill-defined at its location.  This causes trouble for the standard Lorentz force law \eqref{lorentzforce}.  Does our new formulation of the force law \eqref{LCLorentz2} face the same problems?  For a single point charge $q$ at rest, we can use the scalar potential $\phi=\frac{q}{r}$ and set the vector potential to zero.  The force law \eqref{LCLorentz2} yields
\begin{align}
F_i &= - q\left[\left(\frac{\partial}{\partial \eta_i}\right)^p+\left(\frac{\partial}{\partial \xi_i}\right)^p\,\right]\left(\frac{q}{r}\right)
\ .
\end{align}
Focusing on the $x_1$ direction, the first derivative approaches from the left (along the past light-cone) and the second approaches from the right (along the past light-cone).  The two contributions cancel, as they are both infinite but with opposite signs.\footnote{If we take the value of $\phi$ at the particle's location to be positive or negative infinity (depending on the sign of $q$), then the two derivatives are infinite with opposite signs.  If we take the value of $\phi$ at the particle's location to be ill-defined, then the derivatives are ill-defined as well (unless we use Easwaran's open-ended derivatives from footnote \ref{openended1}).}  There is no force on the charge from its own electromagnetic field.  That is nice.  More work could be done to see exactly how well this formulation of electromagnetism can handle self-interaction (and whether it can recover effects like radiation reaction), but it is interesting to see that it has a way to solve the simplest problem of self-interaction.  Perhaps this will suffice for addressing the problems of self-interaction or perhaps we will have to avail ourselves of one of the strategies that have been pursued for solving the problems of self-interaction for point charges in electromagnetism, such as replacing the Lorentz force law with the Lorentz-Dirac force law.\footnote{See \citet[pg.\ 59--63]{frisch2005}; \citet[sec.\ 3]{earman2011}; \citet{Kiessling:2011ab}; \citet[sec.\ 3.1]{lazarovici2018}.}

We have seen that this formulation of electromagnetism with point charges satisfies the first two conditions of the relativistic time evolution paradigm: the dynamical and non-dynamical laws take the correct form.  We have not yet settled whether the formulation satisfies the third condition as we have not yet settled whether the theory is locally deterministic.  To do so, we must figure out whether, in general, an arbitrarily thin law-abiding time slice of a contracting light-cone yields a unique remainder of that light-cone.  This can be proven for the wave equations governing the evolution of the fields,\footnote{See the references in footnote \ref{causalitytheorems}.} but proving it for the entire theory requires looking at the coupled equations for the field and particle dynamics.\footnote{See \citet[sec.\ 2]{frisch2004}; \citet[pg.\ 32--35]{frisch2005}.}  Doing this honestly would require a solution to the aforementioned problems of self-interaction.  There is another complication as well.  \citet{hartensteinhubert} have shown that specifying law-abiding combinations of electromagnetic field and point charge states at an instant generically yields infinities or discontinuities that propagate along the future light-cone (called ``shock fronts'') and leads to ill-defined dynamics.\footnote{See also \citet[sec.\ 8.1]{lazarovici2018}.}  These shock fronts occur because the arbitrarily chosen field around a point charge does not capture the way that the charge's past motion would have acted as a source for the field.  For example, at some moment you might have a particle with an initial velocity surrounded by the Coulomb field of a stationary point charge.  That is not compatible with any reasonable past.  Our formulation of determinism (from section \ref{TEPsection}) requires that we start with an arbitrarily short time interval where the laws are obeyed and that starting point should help in addressing these problems.  We are assuming that the interaction between particle and field has been well-behaved in the immediate past and then asking whether the future evolution is well-behaved and unique.

Shifting now to consider electromagnetism with a continuous charge distribution, the wave equations for the potentials, \eqref{philaw} and \eqref{alaw}, would again be spatiotemporally local dynamical laws.  In addition, we would need law(s) determining the evolution of the charged matter.  The Lorentz force density law \eqref{LCLorentzdensity} can be formulated as an acceptable non-dynamical law, but it will not be a fundamental law for certain classical completions of electromagnetism.  In particular, if we model the evolution of matter using the Klein-Gordon or Dirac equations in Maxwell-Klein-Gordon or Maxwell-Dirac field theory, it is not needed.  As was discussed earlier, the Klein-Gordon equation can be formulated as a spatiotemporally local dynamical law and one might find a way to do so for the Dirac equation.  The Maxwell-Klein-Gordon and Maxwell-Dirac field theories could then satisfy the first two conditions of the relativistic time evolution paradigm.  They should satisfy the third condition as well---the free Klein-Gordon and Dirac equations are locally deterministic\footnote{See \citet[sec.\ 10.1]{wald1984}; \citet[sec.\ 1.5]{thaller1992}; \citet[pg.\ 234, problem 8]{strauss2008}.} and the introduction of interactions does not bring about any infinite or ill-defined self-interaction.\footnote{See \citet{eesr}.}

\section{The Path Ahead}\label{QFTGRsection}

We have thus far focused our attention on parts of physics that seem to fit well with the dynamic production account of laws (Newtonian gravity and electromagnetism), noting the challenges that arise there and exploring how those challenged might be addressed.  Let us now turn briefly to two physical theories that are more hostile to an interpretation in terms of dynamic production: quantum field theory and general relativity.  The purpose of this quick treatment is just to suggest that the dynamic production account has a shot at being extended to our most successful physical theories, and should not be viewed as roadblocked in a way that would make the kind of project pursued here a futile exercise (a worry that was raised at the end of section \ref{TEPsection}).

To build up to quantum field theory, let us start by discussing the non-relativistic quantum mechanics of a fixed number of particles.  In that context, there exist a handful of ``interpretations of quantum mechanics'' that give explicit proposals about the laws and ontology---such as GRW theory, Bohmian mechanics, the many-worlds interpretation, and the many interacting worlds approach.  These four options can all be fit straightforwardly into the time evolution paradigm, though there are other ``retrocausal,'' ``all-at-once,'' or ``non-Markovian'' options\footnote{See \citet{wharton2014, adlam2018, adlam2022, adlam2022b, adlam2023, sep-qm-retrocausality, barandes2023}.} that cannot.  Focusing on the four aforementioned options, the dynamics will be either first-order or second-order depending on the interpretation.  The many-worlds interpretation simply takes the first-order Schr\"{o}dinger equation to give the dynamics of the wave function.  This is naturally interpreted as taking the wave function at a moment as input (with no need to look to the arbitrarily-short past) and returning as output the way the wave function changes into the arbitrarily-short future \citep[pg.\ 857]{easwaran2014}.  GRW theory introduces stochastic exceptions to the Schr\"{o}dinger equation that can be understood as giving probabilities for different behaviors in the arbitrarily-short future \citep[sec.\ 2]{killercollapse}.  Bohmian mechanics is often presented as adding a first-order guidance equation, that is another law beyond the Schr\"{o}dinger equation, determining the velocities of particles from their locations and the wave function (velocities that could be interpreted as future neighborhood properties).  However, a second-order guidance equation can be used instead provided we impose the first-order guidance equation as a restriction on initial conditions \citep{goldstein2015, dewdney2023}.  In the many interacting worlds approach, the dynamics is second-order as particles are reacting to forces by Newton's second law and (at the fundamental level) there is no wave function evolving by the Schr\"{o}dinger equation \citep{HDW, sebens2015, ghadmi2018}.  The fact that quantum theories normally posit first-order dynamics has been used by \citet{builesF} to argue that we live in a world where the present moment is enough to produce the future (the arbitrarily-short past is not needed as input)---in their words, ``our universe is Markovian.''  I prefer to leave open whether the fundamental dynamics will turn out to be first-order or second-order.

Deciding between these competing interpretations of quantum mechanics is difficult.  The many-worlds interpretation has been touted as ``the only game in town'' because it is the only option that can be immediately extended to relativistic quantum field theory (\citealp[pg.\ 35]{wallace2012}; \citealp{wallace2023}).  That virtue could be challenged either because the many-worlds interpretation has other serious problems \citep{adlam2023} or because it is not so easy to formulate quantum field theory in terms of a quantum state evolving by a Schr\"{o}dinger equation \citep{fundamentalityoffields}.\footnote{\citet[sec.\ 2.1]{adlam2022} takes the use of path integrals in quantum field theory to suggest that the theory might best be understood outside the time evolution paradigm.  Although it may be possible to use path integrals to formulate the theory in a way that violates the time evolution paradigm, path integrals in quantum field theory can be viewed as tools for calculating state evolution.}  The problem for Schr\"{o}dinger evolution is that it is unclear whether the quantum state is a particle wave function (in Fock space) or a field wave functional, though hopefully one of these proposals can be made to work.  Although the other interpretations of quantum mechanics may not be as straightforward to extend to quantum field theory as the many-worlds interpretation, there exist a number of promising strategies for doing so.  The attempts to extend GRW and Bohmian mechanics to quantum field theory include non-local interactions that are incompatible with the relativistic time evolution paradigm \citep[ch.\ 7]{maudlin2019}.  This is to be expected, as it is the lesson of EPR and Bell's theorem (taken along with the relevant experimental tests and making certain natural assumptions) that any single-world quantum theory must include such non-local interactions \citep{maudlin2014, maudlin2019}.  One might react to this situation by abandoning the relativistic time evolution paradigm and instead simply seeking theories that fit the original time evolution paradigm, accepting that dynamic production occurs relative to some preferred foliation.  Alternatively, one might work to show that a version of quantum field theory with multiple worlds can be cast in a form that fits the relativistic time evolution paradigm (adopting the many-worlds interpretation or the many interacting worlds approach).  Of course, one could also react to this situation by abandoning dynamic production \citep{adlam2023}.  I simply want to highlight that there are ways to potentially retain dynamic production in quantum field theory.

Let us now move on to general relativity.  In the introduction to their paper, \citet{chengoldstein} present the ``Einstein equation (of general relativity)'' (a.k.a. the Einstein field equations) as one of the motivations for adopting their account of laws, over a dynamic production account, because it is an equation that ``in its usual presentation is non-dynamical.''  Later, they return to the issue and explain that: ``There are ways of converting [the Einstein field equations] into FLOTEs [fundamental laws of temporal evolution] that are suitable for a dynamic productive interpretation.''  \citet{chengoldstein} give the ADM formalism as an example of a way this might be done \citep{arnowitt1962}, complaining that this formalism discards solutions that are not globally hyperbolic (a bug that might be regarded as a feature if our universe is indeed globally hyperbolic and this formalism more precisely narrows down the space of physical possibilities\footnote{See \citet[pg.\ 175, 189--191]{maudlin2007}.}).  Along similar lines, \citet[sec.\ 2.2]{adlam2022} writes
\begin{quote}
``\dots there are a number of theories in modern physics where it doesn't make a great deal of sense to have a rigid division between `state space' and `evolution laws.' For example, a solution to the Einstein equations of General Relativity is not a state at a time but an entire history of a universe \dots\ so it doesn't seem to require any concept of time evolution at all. A time-evolution formulation of the Einstein equations does exist [\citep{ringstrom2009, fouresbruhat1952}], but the original global formulation remains central to research in the field and there seems no obvious reason to think that the time-evolution formulation must be more fundamental.''
\end{quote}
For our purposes here, the most important thing to note is that both \citet{chengoldstein} and \citet{adlam2022} acknowledge that there are versions of general relativity that seem to broadly fit the time evolution paradigm.  So, although there remains work to be done for a proponent of dynamic production, there is a path forward.  One version of general relativity that seems especially well-suited to an interpretation in terms of dynamic production is the ``classical spin-2 field theory'' or ``field-theoretic'' approach (\citealp{salimkhani2020, linnemann2023}; \citealp[sec.\ 5]{gravitationalfield}).  On the field-theoretic approach, gravity is a field on flat Minkowski spacetime, much like the electromagnetic field.  The field equations for this field could potentially be fit into the relativistic time evolution paradigm.  Because the field-theoretic approach treats gravity as similar to other physical interactions, it is arguably an attractive starting point for theories of quantum gravity.

\section{Conclusion}\label{conclusion}

The theories of gravity and electromagnetism that one encounters in introductory physics courses appear to be theories of time evolution that fit well with a dynamic production account of laws of nature.  This paper has sought precision as to how the laws of these theories should be formulated so that they can be understood as producing the future from the past, guiding the ontology in its evolution.  That exercise is important for understanding our options regarding the foundations of these theories and for assessing the merits of the dynamic production account of laws.  The regular and relativistic time evolution paradigms have been put forward to clarify how laws produce the future and we have seen that Newtonian gravity and classical electromagnetism can be fit into these paradigms.  To further develop the dynamic production account, one could build on the preliminary remarks in section \ref{QFTGRsection} to see how more advanced physical theories might be fit into these paradigms.

\section*{Acknowledgements}
I would like to thank Jacob Barandes, Eddy Keming Chen, Chris Dorst, Kenny Easwaran, Jill North, Shelly Yiran Shi, Ward Struyve, Ken Wharton, and anonymous reviewers for helpful comments and discussions.

\end{document}